\documentstyle[12pt]{article}
        \catcode`\@=11
        \@addtoreset{equation}{section}

\def\eq#1{
	(\ref{#1})
}

\catcode`@=11 % This allows us to modify PLAIN macros.
\def\lsim{\mathrel{\mathpalette\@versim<}}
\def\gsim{\mathrel{\mathpalette\@versim>}}
\def\@versim#1#2{\lower0.2ex\vbox{\baselineskip\z@skip\lineskip\z@skip
  \lineskiplimit\z@\ialign{$\m@th#1\hfil##\hfil$\crcr#2\crcr\sim\crcr}}}
\catcode`@=12 % at signs are no longer letters

\begin{document}

%\begin{center}
%{\bf Preliminary Draft}
%\end{center}

\hspace{10.35cm} {\large UICHEP-TH/93-14}

\hspace{12.1cm} {\large DTP/93/68}

\hspace{11.2cm} {\large September 1993}
\vspace*{1cm}
\begin{center}
{\large \bf{Measuring $\Lambda_{\overline{MS}}$ at LEP}}
\end{center}
\vspace*{1cm}
\begin{center}
{\large D.T. Barclay}
\end{center}
%\vspace*{0.25cm}
\begin{center}
{\it Department of Physics, University of Illinois}
\end{center}
\begin{center}
{\it at Chicago,Illinois 60680, U.S.A}
\end{center}
\vspace*{0.5cm}
\begin{center}
{\large C.J. Maxwell and M.T. Reader}
\end{center}
%\vspace*{0.25cm}
\begin{center}
{\it Centre for Particle Theory, University of Durham}
\end{center}
\begin{center}
{\it Durham DH1 3LE, England}
\end{center}
\vspace*{1cm}
\begin{abstract}

We show that standard next-to-leading order (NLO) perturbative QCD
analyses
used to extract $\alpha_{s}$ from LEP data do not serve to disentangle the
completely unknown renormalization scheme (RS) invariant next-NLO
(NNLO) and higher-order uncalculated corrections from
those dependent on the renormalization scale in a predictable manner.
The resulting quoted values of $\alpha_{s}(M_{Z})$
$(\tilde{\Lambda}_{\overline{MS}})$ with attendant scheme dependence
uncertainties therefore reflect the ad hoc way in which they were
extracted, rather than the actual values of these parameters.
Choosing the scale so that the NLO coefficient vanishes (effective
charge scheme) exposes the relative size of these
unknown RS invariant higher order terms. They are seen to be
sizeable for some of the LEP observables, indicating that they must be
estimated if $\tilde{\Lambda}_{\overline{MS}}$ is to be determined
reliably. This can be accomplished either from NNLO calculations, at
present only available for the hadronic width of the $Z^{0}$
($R_{Z}$), or non-perturbatively by writing the RS-invariant
uncertainty in $\tilde{\Lambda}$ in terms of the running
${dR(Q)}/{d \ln Q}$ of the observable R(Q) with energy using an
effective charge formalism. The NNLO calculations for $R_{Z}$ and LEP
data supplemented by lower energy PETRA data lead to
$\tilde{\Lambda}^{(5)}_{\overline{MS}}=287 \pm 100$ MeV. We also
discuss how the effective charge approach can be used to remove scale
dependence from next-to-leading logarithm resummations of some LEP
observables.
\end{abstract}
\renewcommand{\thepage}{}
\vfill
\break
%%%% The introduction
\baselineskip=18pt
\section{Introduction}
\normalsize
\setcounter{page}{1}
\pagenumbering{arabic}

The LEP collaborations now all have high-statistics data samples which
enable them to make accurate measurements of a wide range of
$e^{+}e^{-}$ QCD observables [1-9] - jet fractions,
thrust distributions,
the hadronic width of the $Z^{0}$, energy-energy correlations, etc..
For most of these quantities non-perturbative effects such as
hadronization corrections are expected to be reasonably small and
next-to-leading order (NLO) calculations are available in
renormalization group (RG) improved QCD perturbation theory
\cite{ellis,kunst}. For the
hadronic width of the $Z^{0}$ alone we have a next-NLO (NNLO)
calculation available \cite{gorishny}.

By comparing such calculations with the data one hopes to be able
to extract $\Lambda_{QCD}$, the fundamental SU(3) standard model
parameter, or equivalently $\alpha_{S}$, the $\overline{MS}$ scheme
coupling constant.

Unfortunately this intention is hampered by our rather limited current
knowledge of what QCD actually predicts for any of these quantities due
to the dependence of perturbative calculations at NLO on the unphysical
renormalization scale $\mu$, and more generally on the renormalization
scheme (RS) in higher-orders.
In a complete all-orders calculation of any quantity the $\mu$-dependence
would cancel and so it is an arbitary parameter, but our reliance on a
truncated expansion in $\alpha_{S}$ in which such a cancellation no longer
quite occurs renders predictions ambiguous.

Since it was first clearly formulated in ref. \cite{stevenson} this
`scheme dependence problem' has been the subject of an extensive
literature which has yet to produce consensus.
The response of theorists to the dependence of NLO predictions on
$\mu$ has mainly been to advocate - in the face of the central fact
that it is actually an arbitrary unphysical quantity - that some
particular choice of the scale is the `correct' or `best' one, thereby
removing the ambiguity.
Any such choice must appeal to some criterion outside of the familiar,
well-defined business of calculating Feynman diagrams and no such appeal
has so far been able to command universal support.
Confronted with such confusion amongst their colleagues, experimentalists
have tended to adopt pragmatic compromises that `mix and match' different
proposed solutions.
Specific reservations about the way this has been done at LEP will be
expressed in section 2, but it is clear that any approach to the problem
that increases consistency amongst observables by expanding error bars to
include a `theoretical uncertainty' (essentially obtained by averaging
across camps of theorists) does so by sacrificing unambiguous clarity
over what is being measured, and limits the extent to which QCD could be
investigated in detail.

In some sense, all previously proposed approaches have been an attempt to
second-guess the uncalculated higher-order terms in the perturbation series,
in that they say this choice of scale and a NLO calculation will give
a prediction {\it as if} we knew the full series - and it is likely to be
a reasonable, perhaps good, prediction at that.
On the other hand, our position is essentially that the higher-order terms
are simply unknown and that we therefore should restrict attention to the
more modest aim of deciding what it is actually possible to learn about the
uncalculated terms from the information at our disposal.
A widespread belief already exists that the severity of the scale dependence
for any particular observable is somehow related to the size of its
corrections in higher-orders, although this overlooks the problem that the
size of the unknown terms is itself a scheme dependent notion - there is after
all always a scheme in which there
are no corrections whatsoever.
Put another way, the total uncalculated corrections to a $\mu$-dependent
NLO prediction must also depend on the scale so that their sum does not.
Superficially, we have therefore posed ourselves an unphysical (and hence
uninteresting) question.
However this $\mu$-dependence would only be a fundamental problem if it were
unknown, unpredictable and uncontrollable, when in fact (as we shall show)
the higher-orders can be split into a predictable contribution incorporating
all the awkward $\mu$-dependence and a remaining piece containing all the
genuinely unknown aspects.

This is really just a consequence of the fact that the $\mu$-dependence in
a NLO calculation has a universal form, and so a minimal version of
our claim would be that we are suggesting a simple, convenient means of
comparing theory and data in which the scale dependence is trivial.
The formalism to be proposed has the advantages of being derived
non-perturbatively and as far as possible entirely in terms of RS-invariant
quantities.
In particular, $\alpha_S$ is avoided in favour of $\tilde \Lambda_{\overline{
MS}}$, the dimensional transmutation parameter of QCD, as the free parameter
in the theory since the former is just a theoretical construct without any
direct experimental significance and is indeed strictly a different
quantity in NNLO from what it was in NLO.
It remains true that $\tilde \Lambda$ is RS-dependent, but with the difference
that the values in different schemes are exactly related by only a NLO
calculation.
This $\tilde \Lambda$ is to be as directly related as possible to $R(Q)$, any
generic LEP observable dependent on a single dimensionful scale $Q$
(i.e. $\sqrt s$, the $e^+e^-$ centre of mass energy).
Because this $R(Q)$ will play a role analogous to $\alpha_S$, we shall refer
to this approach as the `effective charge formalism'; it was originally
introduced by Grunberg \cite{grunberg} and has also been discussed by Dhar
and Gupta \cite{dhar}.
Unfortunately, there has been a widespread view that Grunberg's approach
constitutes no more than a particular choice of RS.
Indeed the insistence on using the RS-invariant $R(Q)$ rather than the
RS-dependent $\alpha_S$ in the role of coupling means that the formalism
has much in common with the fastest apparent convergence (FAC) criterion
for choosing the scheme, by which the RS is chosen such that the higher
corrections vanish.
This perturbative property is a nice one, often simplifying calculations,
but in this paper it is entirely secondary to the clarifying benefits of
non-perturbatively identifying each observable as a coupling (from which it
follows trivially).
We cannot overemphasize that the effective charge (EC) formalism is both
derived non-perturbatively and is more general than the adoption of a
particular scheme, even if pieces of that formalism can be interpreted
in terms of the FAC scheme.

Because it should provide a way of measuring $\tilde \Lambda$ without
any scale ambiguity whatsoever, thereby saving the formalism from merely
being a useful framework for displaying data, as important an interpretation
of the formalism derives from the way in which $R(Q)$ runs as a
renormalisation group improved coupling with $Q$.
This evolution is described by a function $\rho(R)$ which is simultaneously
both the basis for a convenient RS-invariant measure of the importance
of the higher-orders and the $\beta$-function corresponding to the FAC scheme.
Thus the effective charge $\beta$-function $\rho(R)$ can, at least in
principle,
be experimentally measured, so that by detailed investigations of the running
of $R(Q)$ with $Q$ one can determine $\tilde \Lambda_{\overline{MS}}$ even in
the absence of NNLO perturbative calculations.
The formalism applied in this way is non-perturbative and transcends the use
of any particular scheme, with the only barrier to an accurate determination
of $\tilde \Lambda_{\overline{MS}}$ being the purely experimental uncertainty.
Of course at present we do not have high
precision information on the running of these LEP observables, and the
running can only be investigated using data obtained at lower energies with
different detectors and machines. In an ideal world LEP would take
data at energies below $\sqrt{s}=M_{Z}$ and study the running of the
observables in detail. As we shall argue, this would enable reasonably
precise determinations of $\tilde{\Lambda}_{\overline{MS}}$, and more
importantly would also indicate whether NNLO corrections by themselves
would  be sufficient, helping to establish whether the extremely
arduous calculations involved are worth embarking on \cite{rellis}.

Even in the absence of information on NNLO corrections and the
detailed running of the observables with energy, however, there is
still information to be obtained with the effective charge formalism,
given only NLO calculations. One can exhibit the relative size of the
uncalculated higher-order corrections for different quantities. In the
effective charge language these are related to how the energy
dependence of the quantity at the experimental energy differs from its
asymptotic dependence as $Q\rightarrow\infty$. If this difference
cannot be neglected, then NNLO calculations or measurements of the
running are necessary before $\tilde{\Lambda}$ can be determined. This
can be decided in the effective charge formalism given only a NLO
calculation and data at one energy.

Our proposal is thus not a solution to the scheme dependence problem in the
conventional sense.
For a start, sufficiently much can be learnt about QCD independently of the
absolute value of $\tilde \Lambda_{\overline{MS}}$ for the problem itself to
be something of a distraction; refining the experimental data on the
relative size of higher-order corrections for different observables and
providing explanations for the aberrant ones is a serious challange in its
own right.
Furthermore, the energy-dependence data already allows us to make tentative
absolute statements in a way complementary to perturbation theory.
The scale dependence problem is not so much solved as completely outflanked.

A preliminary discussion of the effective charge formalism applied in
this way has appeared in references [17] and [18].

The plan of this paper is as follows. In section 2 we shall review the
RS dependence problem and outline the traditional approach to
$\alpha_{S}$ extraction. In section 3 we shall introduce in detail the
effective charge formalism sketched above and stress its advantages
over the conventional approach. We shall attempt to determine
$\tilde{\Lambda}_{\overline{MS}}$ using the rather limited information
on  NNLO corrections and energy dependence of observables currently
available. We shall also discuss an attempt to resum higher-order
terms in the effective charge beta-function using the knowledge
of leading and next-to-leading logarithms in the jet resolution
variable $y_{c}$ for jet rates in the $k_{T}$ or Durham algorithm \cite{bks},
where the perturbation series has an exponentiated structure
\cite{catani,catani2}. Such a
next-to-leading logarithm (NLL) resummation of the effective charge
$\beta$-function is without ambiguity, and in the effective charge
language the $y_{c}$ dependence can be extracted from the experimental
measurements and compared with that of the resummed $\beta$-function,
indicating whether or not the NLL approximation is adequate. Recent
attempts to compare resummed perturbation series with the LEP data require use
of an ad hoc matching procedure \cite{opal1,catani} to include the exact NLO
results, the
additional problem of scale dependence still remaining.

In section 4 we give our conclusions.

%%%%%% Section 2
\section{Review of the Scheme Dependence Problem}

\vspace{1cm}
{\bf A. Parametrizing RS Dependence}
\vspace{1cm}

Consider a generic dimensionless LEP observable $R(Q)$, where $Q$ denotes
the single dimensionful scale on which it depends, typically the
$e^{+}e^{-}$ centre of mass energy $\sqrt{s}$. In RG improved
perturbation theory we can write without loss of generality,
\begin{equation}
\label{RQ}
R(Q) = a + r_{1}a^{2} + r_{2}a^{3} + ....
\end{equation}
where $a$ denotes the RG improved coupling $a\equiv
{\alpha_{s}}/{\pi}$. Notice that by dividing any observable
depending only on a single dimensionful scale by its possibly
dimensionful tree-level perturbative coefficient (which will be RS
invariant), and raising to a suitable power, we can always obtain a
series of the form \eq{RQ} representing a dimensionless $R(Q)$.

The coupling $a$ and series coefficients $r_{i}$ depend on the RS
employed. Using the compact notation and conventions
introduced by Stevenson \cite{stevenson} $a$ is
specified by a $\beta$-function equation
\begin{equation}
\label{beta}
\frac{da}{d\tau}  = -a^{2}(1+ca+c_{2}a^{2}+\ldots)
 \equiv  -\beta(a) ,
\end{equation}
where $\tau\equiv b\ln\frac{\mu}{\tilde{\Lambda}}$, with $\mu$ the
renormalisation scale and $\tilde{\Lambda}$ the dimensional
transmutation mass parameter of QCD, here differing from the
traditional definition by a factor $(2c/b)^{-c/b}$.
$b$ and $c$ are RS invariants
dependent only on the number of quark flavours $N_{f}$ and the number
of colours $N_{c}$; for $N_{c}=3$ we have
$b=(33-2N_{f})/6$ and $c=(153-19N_{f})/2(33-2N_{f})$.
In all our LEP determinations we take $N_f=5$ and assume massless quarks.
For massive quarks the scheme dependence discussion will also go through
in any mass-independent RS, i.e. one in which the coefficients $c_2$, $c_3$
$\ldots$ do not depend on the fermion masses \cite{coquereaux}.

The higher coefficients $c_{2}$, $c_{3}$, $\ldots$ are RS-dependent.
Indeed Stevenson \cite{stevenson}, extending the work of Stueckelberg
and Petermann
\cite{stueckelberg}, has shown that one may
consistently use the parameters ${\tau,c_{2},c_{3},\ldots}$ to label
the renormalization scheme. In the conventional approach when
retaining terms up to and including $r_{n}a^{n+1}$ in equation \eq{RQ}
one truncates the $\beta$-function of equation \eq{beta} retaining terms
up to and including $c_{n}a^{n+2}$. On integrating up the truncated
equation \eq{beta} one can define $a^{(n)}(\tau,c_{2},\ldots,c_{n})$ and
correspondingly one finds for consistency
$r_{1}(\tau)$,$r_{2}(\tau,c_{2})$,$r_{3}(\tau,c_{2},c_{3})$,
$\ldots$,$r_{n}(\tau,c_{2},c_{3},\ldots,c_{n})$.
In this way the $n^{th}$-order truncated approximant is also labelled
by the scheme variables, $R^{(n)}(\tau,c_{2},\ldots,c_{n})$. Of course
when summed to all orders this dependence must cancel and a formally
RS-independent sum be obtained.

The formal consistency of perturbation theory further ensures that
\begin{equation}
R^{(n)}(\tau^{\prime},c_{2}^{\prime},\ldots,c_{n}^{\prime})
- R^{(n)}(\tau,c_{2},\ldots,c_{n})=ka^{n+2} + O(a^{n+3}),
\end{equation}
so that differences between results in two different RS's are formally
effects one order higher in perturbation theory. Of course $k$ depends
on $\{\tau,c_{2},\ldots,c_{n}\}$ and
$\{\tau^{\prime},c_{2}^{\prime},\ldots,c_{n}^{\prime}\}$ and may well not be a
small coefficient.

\vspace{1cm}
{\bf B. The RS dependence of $R^{(1)}(\tau)$.}
\vspace{1cm}

Before discussing the general situation further let us consider the
simplest $n=1$, NLO case. This is the accuracy to which all but one of
the LEP observables have been calculated.

To NLO we have
\begin{equation}
\label{R}
R^{(1)}(\tau) = a^{(1)}(\tau) + r_{1} (\tau)(a^{(1)}(\tau))^{2}
\end{equation}
where $a^{(1)}(\tau)$ is obtained by integrating up the NLO truncation of
equation \eq{beta}
\begin{equation}
\frac{da}{d\tau}=-a^{2}(1+ca).
\end{equation}
With the boundary condition $a^{(1)}(0)=\infty$ one obtains
\begin{equation}
\label{ta}
\tau=\frac{1}{a^{(1)}(\tau)}+c\ln
\left(\frac{ca^{(1)}(\tau)}{1+ca^{(1)}(\tau)} \right)
\equiv F(a^{(1)}(\tau))
\end{equation}
where we define the function $F$ for later use, so that
$a^{(1)}(\tau)=F^{-1}(\tau)$.

How does $r_{1}({\tau})$ depend on $\tau$ explicitly? To see this,
consider two different RS's, $RS$ and $RS^{\prime}$. The connection between
the corresponding renormalised couplings $a$ and $a^{\prime}$ will then be
such that
\begin{equation}
\label{bp}
\beta(a)=\frac{da}{da^{\prime}}\beta^{\prime}(a^{\prime})
\end{equation}
and we can define
\begin{equation}
\label{ap}
a^{\prime}=a(1+\nu_{1}a+\nu_{2}a^{2}+\ldots).
\end{equation}
Inserting \eq{ap} into the series for $R(a^{\prime})$ and $R(a)$ in the two
RS's and equating coefficients one finds
\begin{equation}
\label{r1}
r_{1}=\nu_{1}+r_{1}^{\prime}.
\end{equation}
By integrating up $\beta(a)$ and $\beta'(a')$ as in \eq{ta} and taking
the difference we find
\begin{eqnarray}
\tau-\tau' & = & \frac{1}{a} + c \ln \left(\frac{ca}{1+ca}\right) + O(a)
\nonumber\\
	   &   & -\frac{1}{a'} - c\ln \left(\frac{ca'}{1+ca'}\right) + O(a')
\end{eqnarray}
where the $O(a)$ and $O(a')$ terms reflect contributions beyond NLO in
the $\beta$-functions. Using equation \eq{ap} and equating coefficients
of corresponding powers of $a$ gives
\begin{equation}
\label{tt'}
\tau-\tau'=\nu_{1}.
 \end{equation}
Eliminating $\nu_{1}$ between equations \eq{r1} and \eq{tt'} we
finally arrive at
\begin{equation}
r_{1}'-r_{1}=\tau'-\tau.
\end{equation}
This implies that
\begin{equation}
r_{1}(\tau)=\tau+r_{1}(0).
\end{equation}

It is useful to identify the RS-invariant combination \cite{stevenson}
\begin{equation}
\rho_{0}=\tau-r_{1}(\tau).
\end{equation}
Since $R(Q)$ is a function of a single dimensionful scale $Q$ it follows
that
\begin{equation}
\label{rho0}
\rho_{0}(Q)=\tau-r_{1}(\tau)\equiv b\ln\frac{Q}{\overline{\Lambda}},
\end{equation}
with $\rho_{0}(Q)$ and $\overline{\Lambda}$ RS-invariants.
$\overline{\Lambda}$ is dependent only on the particular QCD
observable $R$. This strongly suggests that these quantities should have
a {\it physical} significance as opposed to RS-dependent quantities
such as $r_{1}(\tau)$ and $a(\tau)$ which depend on unphysical
parameters. As we shall show in section 3 this is indeed the case,
$\rho_{0}(Q)$ and $\overline{\Lambda}$ are connected with the asymptotic
($Q\rightarrow\infty$) dependence of $R(Q)$ on $Q$ with
\begin{equation}
R(Q) {\buildrel Q\to\infty\over\sim}
\frac{1}{\rho_{0}(Q)}=\frac{1}{b\ln\frac{Q}{\overline{\Lambda}}},
\end{equation}
and so $\rho_{0}(Q)$, or equivalently $\overline{\Lambda}$ could
in principle be directly measured given unlimited experimental
energies.

This observable-dependent `physical' quantity $\overline{\Lambda}$ is
directly connected to $\tilde{\Lambda}_{RS}$ the universal dimensional
transmutation parameter of QCD which we are attempting to determine.
To see this, recall that
\begin{equation}
\tau=b\ln\frac{\mu}{\tilde{\Lambda}_{RS}}
\end{equation}
is the variable which specifies the RS in NLO. Notice it is {\it not}
sufficient to specify that one has chosen a given renormalisation
scale $\mu$ in order to specify the RS even at NLO. The
renormalisation scheme is specified by $\tau$ which involves
$\tilde{\Lambda}_{RS}$ as well.

Since $\rho_{0}(Q)$ is an invariant we see that for two different
RS's, $RS$ and $RS'$
\begin{equation}
\label{lnr}
b \ln \frac{\mu}{\tilde{\Lambda}_{RS}}-r_{1}^{RS}(\mu) =
b \ln \frac{\mu}{\tilde{\Lambda}_{RS'}}-r_{1}^{RS'}(\mu),
\end{equation}
where $r_{1}^{RS}(\mu)$ denotes the NLO correction evaluated in the
renormalisation scheme $RS$ and $r_{1}^{RS}(\mu)\equiv r_{1}(\tau)$
with $\tau=b\ln \mu / \tilde \lambda_{RS}$. From equation
\eq{rho0} with $\mu=Q$ we find directly that
\begin{equation}
\label{lam}
\overline{\Lambda}=\tilde{\Lambda}_{RS}\exp(r_{1}^{RS}(\mu=Q)/b)
\end{equation}
and so given a NLO calculation in some RS the invariant
$\overline{\Lambda}$ is exactly related to the dimensional
transmutation parameter $\tilde{\Lambda}_{RS}$. Notice that
$r_{1}^{RS}(\mu=Q)$ is a {\it Q-independent} quantity.

Rearranging equation \eq{lnr} yields
\begin{equation}
\label{Lambda}
\tilde{\Lambda}_{RS'}=\tilde{\Lambda}_{RS}\exp[(r_{1}^{RS}(\mu) -
r_{1}^{RS'}(\mu))/b]
\end{equation}
The exponent is a universal $\mu$-independent constant which exactly
relates the $\tilde{\Lambda}$'s in the two schemes, given NLO
calculations in these two RS's. The implication of this important
result \cite{stevenson,celmaster} is that it does not matter which
$\tilde{\Lambda}_{RS}$ we
choose to try to extract from the data. If the RS is mass-independent
then $\tilde{\Lambda}_{RS}$ will be universal and can be exactly
related to $\tilde{\Lambda}_{RS'}$ in any other scheme by a universal
factor given exactly if we have a NLO calculation for any observable
in both RS's. Thus there is no non-trivial residual scheme dependence
implied in our convenient choice of $\tilde{\Lambda}_{\overline{MS}}$
as the fundamental parameter we wish to extract.

As an illustration consider the minimal subtraction ($MS$)
renormalisation procedure and the $\overline{MS}$ or modified minimal
subtraction procedure where the $\ln(4\pi)-\gamma_{E}$ terms present in
dimensional regularization are subtracted off
$(\gamma_{E}=0.5772\ldots$ is Euler's constant). For any observable,
and independent of $\mu$ \cite{celmaster2},
\begin{equation}
r_{1}^{MS}(\mu) - r_{1}^{\overline{MS}}(\mu) =
\frac{b}{2}(\ln(4\pi)-\gamma_{E}).
\end{equation}
So using equation \eq{lam} we have
\begin{equation}
\tilde{\Lambda}_{\overline{MS}} =
\tilde{\Lambda}_{MS}\exp(\frac{\ln(4\pi)-\gamma_{E}}{2}) =
2.66\tilde{\Lambda}_{MS}.
\end{equation}
This relation is independent of $N_{f}$ or $N_{c}$. For the $MS$ and
$\overline{MS}$ procedures the higher order RS-dependent $\beta$-function
coefficients are identical, $c_{k}^{MS}=c_{k}^{\overline{MS}}$
($k\geq2$). The only difference is in the subtraction procedure and
hence in the definition of the renormalisation scale $\mu$. A choice
of scale $\mu$ using the $MS$ procedure corresponds, therefore, to the
{\it same} renormalization scheme as use of the $\overline{MS}$
procedure with scale $2.66\mu$.

We are finally in a position to exhibit the explicit dependence of
$R^{(1)}(\tau)$ on $\tau$. In fact it is easier to equivalently
consider the dependence on $a^{(1)}(\tau)$. Using equation \eq{rho0}
to write $r_{1}(\tau)$ in terms of $\rho_{0}$ and $\tau$, and using
\eq{ta} for $a^{(1)}(\tau)$ in terms of $\tau$ we find
\begin{equation}
\label{Rt}
R^{(1)}(\tau) = a^{(1)}(\tau) + (a^{(1)}(\tau))^{2}F(a^{(1)}(\tau)) -
\rho_{0}(Q)(a^{(1)}(\tau))^{2}.
\end{equation}
When plotted versus $a(\tau)$, $R^{(1)}$ has the generic approximately
inverted-parabolic shape shown in Figure 1, provided that
$\rho_{0}>0$. From \eq{rho0} this condition will automatically hold
provided that $Q>\overline{\Lambda}$.

\vspace{1cm}
{\bf C. RS dependence of $r_{n}(\tau,c_{2},\ldots,c_{n})$}
\vspace{1cm}

Before we discuss this NLO scheme dependence in more detail let us
complete the picture by discussing the explicit scheme dependence of
the higher coefficients
$r_{2}(\tau,c_{2})$, $r_{3}(\tau,c_{2},c_{3})$,  $\ldots$ as well.
Integrating up the $\beta$-function truncated at $n^{th}$ order with
an infinite constant of integration (related to the definition of
$\tilde{\Lambda}$, see section 3) we obtain, analogous to equation
\eq{ta},
\begin{eqnarray}
\tau & = & \frac{1}{a^{(n)}} + c \ln \left( \frac{ca^{(n)}}{1+ca^{(n)}}
\right) +
\int^{a^{(n)}}_{0} \,dx
\left[ - \frac{1}{x^{2}B^{(n)}(x)}+\frac{1}{x^{2}(1+cx)} \right], \nonumber \\
B^{(n)}(x)&\equiv&(1+cx+c_2x^{2}+c_{3}x^{3}+\ldots+c_{n}x^{n}). \
\end{eqnarray}
This transcendental equation can then be solved explicitly for
$a^{(n)}(\tau,c_{2},\ldots,c_{n})$.

To obtain the explicit RS dependence of the $r_{i}$ it is
convenient to identify a special RS, the effective charge (EC) scheme
\cite{grunberg,dhar},
in which $r_{1}=r_{2}=r_{3}=\ldots=r_{n}=0$. Then $R^{(n)}=a^{(n)}$ in
this scheme, and the coupling constant is the physical observable.
This choice of RS will correspond to particular values of the scheme
parameters,
$\tau_{EC}=\rho_{0},c_{2}^{EC}\equiv
\rho_{2},c_{3}^{EC}\equiv\rho_{3} \ldots$.

The fact that $\tau_{EC}=\rho_{0}$ follows directly from equation
\eq{rho0},
$r_{1}(\tau)=\tau-\rho_{0}$
then
$r_{1}(\tau_{EC})=\tau_{EC}-\rho_{0}=0~~\Rightarrow~~\tau_{EC}=\rho_{0}$.
 From equation \eq{bp} we have that for two RS's, barred and
unbarred,
\begin{equation}
\label{betabar}
\overline{\beta}(\overline{a})=\frac{d\overline{a}}{da}\beta(a(\overline{a})).
\end{equation}
If the barred RS is chosen to be the EC scheme then
\begin{equation}
\label{betabar2}
\overline{\beta}(\overline{a})\equiv\rho(\overline{a}) =
\overline{a}^{2}(1 + c\overline{a} + \rho_{2}\overline{a}^{2} +
\ldots+\rho_{k}\overline{a}^{k} + \ldots),
\end{equation}
with $\overline{a}=R$. Then \eq{betabar} gives
\begin{equation}
\rho(R) = \beta(a(R))\frac{dR}{da},
\end{equation}
where $a(R)$ is the inverted perturbation series. By expanding out the
right-hand side as a power series in $R$ and equating coefficients we
obtain \cite{C.J.Maxwell}
\begin{eqnarray}
\label{rho2}
\rho_{2} & = & c_{2} + r_{2} - r_{1}c - r_{1}^{2} \nonumber \\
\rho_{3} & = & c_{3} + 2r_{3}
- 4r_{1}r_{2} - 2r_{1}\rho_{2} -r_{1}^{2}c + 2r_{1}^{3} \\
\vdots & & \vdots \nonumber
\end{eqnarray}
These objects are $Q$-independent RS invariants constructed from
the $r_{i}$ and $c_{i}$ in any RS. For instance given NLO and NNLO
calculations of $r_{1}^{\overline{MS}}(Q),r_{2}^{\overline{MS}}(Q)$ and
since $c_{2}^{\overline{MS}}$ is known \cite{tarasov,larin}, one can construct
\begin{equation}
\rho_{2} = c_{2}^{\overline{MS}} + r_{2}^{\overline{MS}} -
r_{1}^{\overline{MS}}c - (r_{1}^{\overline{MS}})^{2}.
\end{equation}
As we shall see in the next section just like the NLO invariant
$\rho_{0}$, the NNLO and higher $\rho_{2},\rho_{3},\ldots$ invariants
have a physical significance, whereas quantities such as
$r_{1}^{\overline{MS}}(Q),c_{2}^{\overline{MS}}$ are intermediate
RS-dependent quantities which should be eventually eliminated in favour
of RS invariants.

Having obtained $\rho_{2},\rho_{3}$ from NNLO, and higher order
calculations in any convenient RS, we can exhibit the explicit
$\tau,c_{2},c_{3},\ldots$ dependence of the perturbative
coefficients by rearranging equation \eq{rho2}. We have that
\begin{eqnarray}
\label{list}
r_{1}(\tau) & = & (\tau - \rho_{0}), \nonumber \\
r_{2}(\tau,c_{2}) & = & (\tau - \rho_{0})^{2} + c(\tau - \rho_{0}) +
(\rho_{2}-c_{2}), \nonumber \\
r_{3}(\tau,c_{2},c_{3}) & = & (\tau - \rho_{0})^{3} +
\frac{5}{2}c(\tau - \rho_{0})^2 + (3\rho_{2}-2c_{2})(\tau-\rho_{0}) +
\frac{1}{2}(\rho_{3}-c_{3}) \\
\vdots & & \vdots \nonumber
\end{eqnarray}
The result for $r_{n}(\tau,c_{2},\ldots,c_{n})$ is a polynomial in
$(\tau-\rho_{0})$ of degree $n$ with coefficients involving
$\rho_{n},\rho_{n-1},\ldots,c$ and $c_{2},c_{3},\ldots,c_{n}$, such
that $r_{n}(\rho_{0},\rho_{2},\ldots,\rho_{n})=0$.

\vspace{1cm}
{\bf D. NLO extraction of $\alpha_{s}(M_{Z})$.}
\vspace{1cm}

Let us now return to Figure 1 and discuss how one might attempt to use
it to extract $\tilde{\Lambda}_{\overline{MS}}$ or $\alpha_{s}(M_{Z})$
in the $\overline{MS}$ scheme.

Notice first that the curve $R^{(1)}(\tau)$ of equation \eq{Rt} is a
{\it universal} function of $\tau$ and $\rho_{0}$ (for given fixed
$N_{f}$), where the value of
the invariant $\rho_{0}$ depends on the particular observable $R$.
$\rho_{0}$ is of course directly related to $\tilde{\Lambda}_{\overline{MS}}$
given a NLO calculation of $r_{1}^{\overline{MS}}(Q)$, with
\begin{equation}
\label{rho}
\rho_{0}=b\ln\frac{Q}{\tilde{\Lambda}_{\overline{MS}}} -
r_{1}^{\overline{MS}}(Q).
\end{equation}
$R^{(1)}(\tau)$ has a maximum at $a(\tau)\simeq 1/\rho_{0}$
($\tau\simeq\rho_{0}$) where
$R^{(1)}(\rho_{0})\simeq 1/\rho_{0}$. The full-width at half-maximum
in $a(\tau)$ is approximately $\sqrt{2}/\rho_{0}$. $R^{(1)}$
vanishes at $a=0$ ($\tau=\infty$) and at $a(\tau) \simeq
2/ \rho_0$ ($\tau \simeq \rho_{0}/2$). For comparison Figure
1 is drawn with $\rho_{0}=25$ and $N_{f}=5$.

Let us suppose that the horizontal dashed line in Figure 1
represents the measured experimental data. If $\rho_{0}$ is
sufficiently small that the maximum
$R^{(1)}\simeq 1 / \rho_{0}$
lies above the data then the curve will cut the data at two scales
$\tau_{1},\tau_{2}$ (as
illustrated). Conversely, if $\rho_{0}$ is made larger the curve will
be below the data line. Thus an infinite set of $\tau,\rho_{0}$ pairs
fit the data perfectly. If we wish to {\it measure} $\rho_{0}$ (and
hence $\tilde{\Lambda}_{\overline{MS}}$) we must {\it specify} $\tau$.
This is the NLO scheme dependence problem. At NNLO one would have a
surface $R^{(2)}(\tau,c_{2})$ and to extract $\rho_{0}$ one would need
to specify $\tau$ and $c_{2}$, and so on with one extra unphysical
parameter for each order in perturbation theory. At least in NLO for a
given value of $\rho_{0}$ ($\tilde{\Lambda}_{\overline{MS}}$) there is a
{\it maximum} possible $R^{(1)}$.

In NNLO and higher one can show \cite{thesis} that for a given value of
$\rho_{0}$
($\tilde{\Lambda}_{\overline{MS}}$) there exists a choice of
$\tau,c_{2}$, $\ldots$, $c_{n}$ such that $R^{(n)}$ has {\it any} desired
positive value. Thus, for {\it any} $\tilde{\Lambda}_{\overline{MS}}$
we can choose a sequence of schemes such that
$R^{(2)}=R^{(3)}=\ldots=R^{(n)}=R_{exp}$, the experimentally measured
value.

Various ``solutions'' of the scheme dependence problem, i.e.
motivations for particular choices of $\tau,c_{2},\ldots,c_{n}$ have
been proposed.

\vspace{1cm}
(1) {\it Principle of Minimal Sensitivity (PMS)} \cite{stevenson}
\vspace{1cm}

The idea here is that since the exact (all-orders) result is
independent of $\tau$,$c_{2}$,$c_{3}$,$\ldots c_{n}$ one should choose
the $n^{th}$-order approximation $R^{(n)}(\tau,c_{2},\ldots,c_{c})$ to
mimic this property and be as insensitive as possible to the chosen
value of the unphysical parameters. That is one arranges that
\begin{equation}
\left. \frac{\partial R^{(n)}}{\partial \tau} \right|_{\tau=\overline{\tau}} =
\left. \frac{\partial R^{(n)}}{\partial c_{2}}
\right|_{c_{2}=\overline{c}_{2}} = \ldots = \left. \frac{\partial
R^{(n)}}{\partial c_{n}} \right|_{c_{n}=\overline{c}_{n}}=0
\end{equation}
and the PMS scheme is specified by
$\overline{\tau},\overline{c}_{2},\ldots,\overline{c}_{n}$. At NLO one
has to solve
\begin{equation}
\left. \frac{\partial R^{(1)}}{\partial \tau}
\right|_{\tau=\overline{\tau}} = 0,
\end{equation}
and this corresponds to solving the transcendental equation
\begin{equation}
\frac{1}{a} + c \ln \left[ \frac{ca}{1+ca} \right] +
\frac{1}{2}\frac{c}{1+ca} = \rho_{0}.
\end{equation}
The solution $a(\overline{\tau})=\overline{a}$ then gives
\begin{equation}
R^{(1)}_{PMS} =
\frac{\overline{a}(1+\frac{1}{2}c\overline{a})}{(1+c\overline{a})}.
\end{equation}
The stationary PMS point of $R^{(1)}(\tau)$ is the maximum at
$\tau=\overline{\tau} \simeq \rho_{0}-c/2$ (this is a somewhat
better approximation than the cruder $\tau \simeq \rho_{0}$ given
previously for the position of the maximum). $\rho_{0}$ is adjusted so
that $R^{(1)}_{PMS}=R_{exp}$.

\vspace{1cm}
(2) {\it Effective Charge (EC) Scheme} \cite{grunberg,dhar}
\vspace{1cm}

This corresponds to a choice of scheme such that
$r_{1}=r_{2}=\ldots=r_{n}=0$, hence $a^{(n)}=R^{(n)}$ is an effective
charge. As we have seen the scheme parameters are then
$\{\rho_{0},\rho_{2},\rho_{3},\ldots,\rho_{n} \}$, where the
$\rho_{i}$ are the RS-invariants in equation \eq{rho2}. Once these
have been computed up to $\rho_{n}$ from higher-order calculations in
any technically-convenient RS (e.g. $\overline{MS}$) they can be
inserted in the integrated $\beta$-function equation (2.24) which
can then be solved with $a^{(n)}=R$ for given $\rho_{0}$. $\rho_{0}$ can
then be obtained by requiring $R=R_{exp}$.

This particular approach is sometimes also referred to as the fastest
apparent convergence criterion. In this language it appears rather
artificial, but as we shall emphasise in section 3 its advantage is
that the all-orders coupling constant {\it and} $\beta$-function are
experimentally observable physical quantities, allowing a {\it
non-perturbative} approach to supplement
$\tilde{\Lambda}_{\overline{MS}}$ measurement, and a physical
definition of the uncertainty.

It should be emphasised that EC and PMS predictions in NLO are very
similar since, as we have seen, $\tau_{EC}=\rho_{0}$, $\tau_{PMS} \simeq
\rho_{0} - c/2$. It is also true that in NNLO EC and PMS
remain close to each other \cite{nicholls,chyla}. For the PMS approach,
however, the
coupling constant and $\beta$-function are unphysical quantities, and it
is not clear even that their all-orders versions are defined, given
the complex nature of the coupled equations which must be solved.

\vspace{1cm}
(3) {\it Physical Scale} \cite{thephys}
\vspace{1cm}

According to this viewpoint the renormalization scale should be chosen
to be close to
the physical scale in the problem, $\mu=Q$. If predictions in the
vicinity of $\mu=Q$ are strongly $\mu$-dependent then this is
supposedly an indication that the perturbation series is
intrinsically badly behaved. This viewpoint is usually justified by
noting that perturbative coefficients in higher orders will be
polynomials in $\ln \mu/ Q$,
\begin{equation}
\label{rn}
r_{n}=\sum_{l=0}^{n}K_{nl}(b\ln\frac{\mu}{Q})^{l}.
\end{equation}
To avoid unnecessarily large logarithms one should therefore set
$\mu\simeq Q$. Whilst this is true as far as it goes, it tacitly
assumes that the NLO RS dependence is completely given by the
dependence on the renormalisation scale. In fact as we have seen
$r_{n}(\tau,c_{2},\ldots,c_{n})$ where
$\tau=b\ln \mu/\tilde \Lambda_{RS}$ so the coefficients $K_{nl}$
in \eq{rn} will depend on $\tilde{\Lambda}_{RS}$. The meaning of
$\mu$ of course depends for instance on whether it is
$\mu(\overline{MS})$ or $\mu(MS)$. In contrast we have from \eq{list} that
\begin{equation}
\label{rnt}
r_{n}(\tau,c_{2},\ldots,c_{n})=\sum_{l=0}^{n}\tilde{K}_{nl}(\tau-\rho_{0})^{l}.
\end{equation}
So to avoid unnecessarily large terms we should clearly choose
$\tau\simeq \rho_{0}$, the effective charge scheme! To make contact
with \eq{rn} notice that
\begin{equation}
(\tau-\rho_{0}) = \left[ b\ln\frac{\mu}{Q} + r_{1}^{RS}(Q) \right]
= b \ln \left[ \frac{\mu}{Q} e^{r_{1}^{RS}(Q)/b} \right] .
\end{equation}
So we can write \eq{rnt} as
\begin{equation}
\label{rntc}
r_{n}(\tau,c_{2},\ldots,c_{n})=\sum_{l=0}^{n}\tilde{K}_{nl} \left[
b\ln \left( \frac{\mu e^{r_{1}^{RS}(Q)/b}}{Q} \right ) \right]^{l}
\end{equation}
In \eq{rnt} and \eq{rntc} the coefficients $\tilde{K}_{nl}$ do not
depend on the NLO RS choice, only on $c_{2},c_{3},\ldots$ and the RS
invariants $\rho_{2},\rho_{3},\ldots$, whereas in \eq{rn} the $K_{nl}$
will depend on $r_{1}^{RS}(Q)$ and so have a hidden dependence on the
NLO RS choice which is customarily ignored in the usual `physical
scale' argument. Applying the argument to \eq{rntc} instead one would
infer that one should set $\mu \simeq Q e^{-r_{1}^{RS}(Q)/b} = Q
\tilde \Lambda_{RS} /{\overline{\Lambda}} $, which of course
corresponds to the effective charge scheme, $\tau=\rho_{0}$.
For particular cases the extra factor may be large.

Recall that $\overline{\Lambda}$ is an RS-invariant characterising how
the observable runs with $Q$ asymptotically. Different observables will
have different $\overline{\Lambda}$'s. If it happens that
$\tilde{\Lambda}_{RS} \simeq \overline{\Lambda}$ ($r_{1}^{RS}(Q)
\simeq 0$) then predictions will be stable and large logarithms
avoided for $\mu \simeq Q$, but no special physical significance should
be accorded to this fact. It is always possible to modify the
subtraction procedure so that
$\tilde{\Lambda}_{RS}=\overline{\Lambda}$, since $\tilde{\Lambda}$'s
in different RS's are related according to \eq{Lambda}.

We conclude that a modified `avoidance of unnecessarily
large logarithms' physical scale argument which correctly labels the
NLO RS dependence actually leads to the effective charge scheme,
$\tau=\rho_{0}$.

\vspace{1cm}
{\it (4) Fitting $\mu$ and $\tilde{\Lambda}$ to the data}
\vspace{1cm}

For most of the LEP observables we have a differential distribution in
a kinematical variable rather than a single experimental data value as
in Figure 1. For instance a thrust distribution in the thrust variable
$T$, or a multijet rate in the jet resolution cut $y_{c}$. Denoting such
a generic kinematical variable by $\lambda$, we should really add an
extra perpendicular axis to Figure 1 and consider the $\lambda$
dependence as well, $R^{(1)}(\tau,\lambda)$. The NLO coefficient will
also depend on $\lambda$ and so
for a given choice of $\tilde{\Lambda}_{\overline{MS}}$ we will
have $\rho_{0}(\lambda)$ from \eq{rho}. If
$\tilde{\Lambda}_{\overline{MS}}$ is sufficiently large then the
experimental data $R_{exp}(\lambda)$ will intersect
$R^{(1)}(\tau,\lambda)$ for at least one $\tau$ value for any
$\lambda$ and so the data can be fitted perfectly with a suitable
$\lambda$-dependent RS parameter choice $\tau(\lambda)$. In general
there will be two intersection points as in Figure 1 and so two
functions $\tau_{1}(\lambda),\tau_{2}(\lambda)$ will fit the data
perfectly. Since this can be done for any suitably large
$\tilde{\Lambda}_{\overline{MS}}$ there is no unique best fit.

The OPAL analysis \cite{opal1,opal2} uses a mixture of approaches (3) and (4).
The data
for each observable over a range of the corresponding kinematical
variable is compared with the NLO prediction with the
$\overline{MS}$ scale chosen as $\mu=M_{Z}$, and a best fit for
$\tilde{\Lambda}_{\overline{MS}}$ performed. The resulting
$\alpha_{S}(M_{Z})$ values are shown in Figure 2(a), reproduced from
reference \cite{sbethke}. There is evidently a considerable scatter in the
$\alpha_{S}$ values obtained.

One then performs a simultaneous best fit for $\mu$ and
$\tilde{\Lambda}_{\overline{MS}}$ for each observable over a range of
the kinematical variables. This provides a second
$\alpha_{S}(M_{Z})$ value. The quoted central value of
$\alpha_{S}(M_{Z})$ is then taken to lie mid-way between these
extremes (perhaps modified by hadronization) and the difference between
them is taken as indicative of the
size of the uncalculated higher-order corrections. By enlarging the
`scheme dependence error' in this way one obtains greater apparent
consistency between different observables as shown in Figure 2(b).

In our view one learns nothing useful from an analysis of this kind.
Consider Figure 2(a) and the considerable scatter of $\alpha_{S}$ values
obtained from different observables fitted with $\mu=M_{Z}$. Choosing
$\mu=M_{Z}$ means that from \eq{rntc} the higher-order coefficients are
given by
\begin{equation}
\label{rnu}
r_{n}(\mu=M_{Z})=\sum_{l=0}^{n}
\tilde{K}_{nl}(r_{1}^{\overline{MS}}(M_{Z}))^{l}
\end{equation}
where the $\tilde{K}_{nl}$ depend on the invariants
$\rho_{2},\rho_{3},\ldots,\rho_{n}$ and on the NNLO and higher RS
parameters $c_{2},c_{3},\ldots,c_{n}$. We have tabulated in Table 1
the $r_{1}^{\overline{MS}}(M_{Z})$ values for some of the observables at
particular values of the kinematical parameters using the NLO matrix
elements of reference 10. (For more details see section 3B).
One sees that they are large and rather
scattered, e.g for the jet rates
$r_{1}^{\overline{MS}}(M_{Z}) \sim 10$ which results in rather small EC (or
PMS) scales, $\mu_{EC} \sim 0.07M_{Z}$. As emphasized earlier no
physical significance attaches to these small scales, since the scale
itself is just an artifact of the unphysical subtraction procedure
employed. Thus from \eq{rnu} one may expect large and rather scattered
higher order corrections for the different observables with
$\mu=M_{Z}$, which translates itself into the expectation of scattered
values of $\alpha_{S}(M_{Z})$, which indeed are observed.

If we instead choose $\mu=M_{Z} e^{-r_{1}^{\overline{MS}}(M_{Z})/b}
\equiv \mu_{EC}$ (the effective charge scale) for each observable then
the higher order coefficients are given by
\begin{equation}
r_{n}(\mu=\mu_{EC}) = \tilde{K}_{n0}
\end{equation}
where $\tilde{K}_{n0}$ does not depend on the NLO RS choice ($\tau$),
only on the NNLO and higher RS parameters and the RS-invariants
$\rho_{2},\rho_{3},\ldots$. Any scatter now observed in the extracted
$\tilde{\Lambda}_{\overline{MS}}$ or $\alpha_{S}(M_{Z})$ for different
observables will be attributable to these NNLO and higher RS
invariants being large, and not to the scatter due to the NLO scheme
dependence logarithms involving $r_{1}^{\overline{MS}}(M_{Z})$, which
are avoided by the choice $\mu=\mu_{EC}$. As we shall discuss in
Section 3 there is still scatter when this is done but it is somewhat
reduced.

With the choice $\mu=M_{Z}$ predictable scatter due to {\it
avoidable} already known NLO scheme dependence logarithms is
superimposed on top of the interesting scatter due to the size of the
NNLO and higher
invariants which is telling us about NNLO and higher order uncalculated
corrections. The choice $\mu=\mu_{EC}$ removes the predictable scatter
and provides genuine information on the interesting NNLO and
higher-order uncalculated corrections.

The $\overline{MS}$ scales obtained by fitting to the data for an
observable over some range of the associated kinematical parameters
tend to be considerably smaller than $M_{Z}$, as for the EC or PMS
scale. Typically for values of the kinematical variable $\lambda$ well
away from the 2-jet region where large logarithms will be important,
$\rho_{0}(\lambda)$ and $r_{1}(\lambda)$ do not vary strongly with
$\lambda$. This means that one cannot obtain best fits over such a
range of $\lambda$. It is only by including the 2-jet region that one
can obtain stable best fits, but then fitting for a
$\lambda$-independent scale over a range where $\rho_{0}(\lambda)$ and
$r_{1}(\lambda)$ (and presumably uncalculated higher order
corrections) are strongly $\lambda$-dependent represents a rather
severe compromise, and it is not obvious what one learns.

In conclusion, we believe that the standard NLO analyses used to
extract $\alpha_{S}$ from LEP data do not serve to disentangle the
genuinely new and interesting NNLO and higher order uncalculated
corrections from predictable higher order corrections connected with
the choice of RS at NLO. The resulting quoted values of
$\alpha_{S}(M_{Z})$ (or $\tilde{\Lambda}_{\overline{MS}}$) with attendant
scheme dependence uncertainties therefore reflect the ad hoc way in
which they were extracted, rather than the actual values of these
parameters. In contrast, by choosing the effective charge scale for
each observable one exposes the relative importance of uncalculated
NNLO and higher corrections. An effective charge formalism which can
be supplemented with non-perturbative information is described in the
next section.

 %%%%Section 3
\vspace{0.5cm}
\section{The Effective Charge Formalism}
\vspace{.5cm}

{\bf A. The $Q$ dependence of $R(Q)$}
\vspace{1cm}

For the dimensionless QCD observable $R(Q)$ we can define
\begin{equation}
\label{zeta}
\frac{dR}{d \ln Q} \equiv \xi(R).
\end{equation}
${dR}/{d \ln Q}$ and hence $\xi(R(Q))$ are, at least
in principle, experimentally observable quantities, although collider
experiments are usually designed to make high-statistics observations
at a fixed energy $Q$ rather than examining the detailed running of
observables with energy. To make contact with the standard
perturbative approach we note that \eq{zeta} is the $\beta$-function
equation in the effective charge scheme where the coupling $a=R(Q)$
and the $\beta$-function is $\rho(R)$ as defined in equation \eq{betabar2}. So
we have
%\begin{eqnarray}
%\label{dR}
%\frac{dR}{d \ln Q} & = & \xi(R(Q)) = -b \rho(R) \nonumber \\
%& \approx & -b R^{2}(1 + cR + \rho_{2}R^{2} + \ldots) +
%e^{-S/R}R^{\delta}(K_{0}+K_{1}R+\ldots)
%\end{eqnarray}
\begin{equation}
\label{dR}
\frac{dR}{d \ln Q}= \xi(R(Q)) = -b \rho(R) \approx
-b R^{2}(1 + cR + \rho_{2}R^{2} + \ldots) +
e^{-S/R}R^{\delta}(K_{0}+K_{1}R+\ldots).
\end{equation}
The effective charge $\beta$-function $\rho(R)$ may be regarded as a
physical observable. As measured from data it will include a resummed
version of the (perhaps asymptotic) formal perturbative series
exhibited in \eq{dR} together with the non-perturbative terms involving
$e^{-1/R}$ which are invisible in perturbation theory. (Here
S,$\delta$,$K_{0}$,$K_{1}$,\ldots are observable-dependent constants
which we shall not specify further \cite{beneke}.) Hence by
comparing with data one can test how well the first few terms of the
perturbative series serve to represent the observed running of $R(Q)$.
That is, from measurements of $R_{exp}(Q)$ and $dR_{exp}(Q)/d \ln Q$,
one can test the extent to which
\begin{equation}
\frac{dR_{exp}}{d \ln Q} \approx -b R_{exp}^{2}(1+cR_{exp})
\end{equation}
describes the data. A marked discrepancy would either indicate
the importance of the NNLO invariant $\rho_{2}$ and higher terms, or
the relative importance of the non-perturbative contributions, or both.

Given a collider program dedicated to the investigation of the
$Q$-dependence of observables the above tests would indeed be powerful
and useful. The focus of present studies, however, is the accurate
measurement of $R$ at fixed $Q$, i.e. $Q=M_{Z}$, and the comparison of
these results with NLO matrix element calculations in order to extract
$\tilde{\Lambda}_{\overline{MS}}$. To obtain $R(Q)$ we clearly need to
integrate equation \eq{zeta}. The boundary condition will be the assumption of
asymptotic freedom, that is $R(Q) \rightarrow 0$ as $Q \rightarrow
\infty$, which corresponds to the requirement that $\xi(R(Q)) < 0$
for $Q>Q_{0}$, with $Q_{0}$ some suitably low energy. (Equivalently,
$\rho(R(Q))>0$ for $Q>Q_{0}$, assuming $b>0$ or $N_{f}< 33/2$ for
$N_{C}=3$ QCD.) Any zero of $\xi(R(Q))$, $\xi(R^{*})=0$, say,
would correspond to $R(Q) \rightarrow R^{*}$ as $Q \rightarrow \infty$
and ultraviolet fixed point behaviour.

Integrating equation \eq{zeta} we obtain
\begin{equation}
\label{lnQL}
\ln \frac{Q}{\overline{\Lambda}} = \int^{R(Q)}_{0} \frac{1}{\xi(x)}dx
+ (infinite~constant)
\end{equation}
where $\overline{\Lambda}$ is a finite constant which depends on the
way the infinite constant is chosen. The infinite constant can be
chosen to be
\begin{equation}
\label{constant}
(Infinite~constant)=-\int^{\infty}_{0} \frac{dx}{\eta(x)},
\end{equation}
where $\eta(x)$ is {\it any} function which has the same $ x
\rightarrow 0$ behaviour as $\xi(x)$. We know from equation \eq{dR}
that $\xi(x)$ has the universal $x \rightarrow 0$ behaviour
$\xi(x)=-b \rho (x) \approx -bx^{2}(1+cx)$ so we choose
$\eta(x)=-bx^{2}(1+cx)$. Inserting this choice for $\eta(x)$ and
rearranging equation \eq{lnQL} we find
\begin{equation}
b \ln
\frac{Q}{\overline{\Lambda}}=\int^{\infty}_{R(Q)}\frac{dx}{x^{2}(1+cx)}
+ \int^{R(Q)}_{0}dx \left[ -\frac{1}{\rho(x)}+\frac{1}{x^{2}(1+cx)}
\right].
\end{equation}
The first integral on the right-hand side is just $F(R)$ where $F$ is the
function
defined in equation (2.6),
\begin{equation}
F(x) \equiv \frac{1}{x} + c \ln \left[ \frac{cx}{1+cx} \right].
\end{equation}

We define for later convenience
\begin{equation}
\label{deltarho}
\Delta\rho_{0}(Q) \equiv \int^{R(Q)}_{0} dx \left[
-\frac{1}{\rho(x)}+\frac{1}{x^{2}(1+cx)}
\right].
\end{equation}
Notice that despite appearances the integrand of $\Delta \rho_{0}$ is
regular at $x=0$. Then we have
\begin{equation}
\label{FRQ}
F(R(Q)) = b \ln \frac{Q}{\overline{\Lambda}} - \Delta\rho_{0}(Q).
\end{equation}
As $Q \rightarrow \infty$, $R(Q) \rightarrow 0$ so that $\Delta
\rho_{0} \rightarrow 0$ and thus asymptotically
\begin{equation}
F(R(Q)) \approx b \ln \frac{Q}{\overline{\Lambda}}
\end{equation}
with the constant of integration given by
\begin{equation}
\label{Lbar}
\overline{\Lambda}= \lim_{Q \rightarrow \infty} Q \exp(-F(R(Q))/b).
\end{equation}
Notice that in arriving at (3.9) we did not need to refer to
perturbation theory, except to assume the asymptotic $x \rightarrow 0$
$(Q \rightarrow \infty)$ behaviour $\xi(R(Q)) \approx
-bR^{2}(1+cR)$.
Despite this, the constant $\overline{\Lambda}$ obtained with the
particular choice $\eta(x)=-bx^{2}(1+cx)$ is precisely the
$\overline{\Lambda}$ introduced in equation \eq{rho0}. That is
\begin{equation}
\label{Lbara}
\overline{\Lambda}= \tilde{\Lambda}_{\overline{MS}}
\exp(r_{1}^{\overline{MS}}(\mu=Q)/b).
\end{equation}
We can see this immediately since if $a$ denotes the coupling in the
RS with $\mu = Q$ using the $\overline{MS}$ subtraction procedure in
NLO, and with higher order $\beta$-function coefficients zero \cite{hooft} we
have
\begin{equation}
\label{blnQ}
b \ln \frac{Q}{\tilde{\Lambda}_{\overline{MS}}} = F(a)
\end{equation}
giving a well-defined all-orders coupling. Further
\begin{equation}
R=a+r_{1}^{\overline{MS}}(Q)a^{2}+\ldots.
\end{equation}
$a \rightarrow 0$ as $Q \rightarrow \infty$, so asymptotically as $Q
\rightarrow \infty$ we have
\begin{equation}
\label{F(R)}
F(R) \approx F(a) - r_{1}^{\overline{MS}}(Q) + \ldots
\end{equation}
where the ellipsis denotes terms which vanish as $Q \rightarrow
\infty$. Inserting \eq{F(R)} into \eq{Lbar} and using \eq{blnQ} we
find \eq{Lbara} as $Q \rightarrow \infty$.

So finally we have
\begin{eqnarray}
\label{FRQ}
F(R(Q)) & = & b \ln \frac{Q}{\overline{\Lambda}} - \Delta \rho_{0}(Q)
\nonumber \\
& = & b \ln \frac{Q}{\tilde{\Lambda}_{\overline{MS}}} -
r_{1}^{\overline{MS}}(Q) - \Delta \rho_{0}(Q) \\
& =& \rho_{0}(Q) - \Delta\rho_{0}(Q). \nonumber
\end{eqnarray}
We could, of course, have written down the result of equation \eq{FRQ}
perturbatively in the EC scheme at once by simply using the integrated
beta-function equation of equation (2.24) with $\tau =
\rho_{0}$, $B^{(n)} = \rho^{(n)} =
1+cx+\rho_{2}x^{2}+\ldots+\rho_{n}x^{n}$, and $a^{(n)}=R^{(n)}$. However, our
purpose here has been to stress that equation \eq{FRQ} holds {\it
beyond perturbation theory} for the measured observable $R(Q)$ and
$\Delta \rho_{0} (Q)$ constructed from the measured running of $R(Q)$,
${dR(Q)}/{d \ln Q}=-b\rho(R(Q))$. That is, we can write a {\it
non-perturbative} closed expression {\it exactly} relating
the universal constant $\tilde{\Lambda}_{\overline{MS}}$ to
observable quantites.

Since $\Delta \rho_{0}(Q_{0})$ involves an integral in $R(Q)$ between
$R(Q_{0})$ and 0, to actually measure it would require knowlege of $R(Q)$
(equivalently ${dR}/{d \ln Q}$) on the full range $[Q_{0},\infty)$. We of
course
only know $R(Q)$ on some finite energy range, so if we want to obtain
$\tilde{\Lambda}_{\overline{MS}}$ from equation \eq{FRQ} and
measurements of $R(Q)$ there will always be an uncertainty related to
our lack of knowledge about the behaviour of $R(Q)$ beyond the highest
energy reached, or correspondingly related to our lack of knowledge
about $\rho(R)$ for $R<R(Q_{0})$, with $Q_{0}$ the highest energy
reached. As we shall discuss in sub-section 3C we can use measurements
of the running of $R(Q)$ to determine $\rho(R)$ in the vicinity of
$R=R(Q_{0})$, and perturbative QCD calculations to determine $\rho(R)$
in the vicinity of $R=0$. Putting these two pieces of information
together we can make unambiguous statements about the validity of
perturbation theory and the nature of the function $\rho(R)$, which
can assist our attempts to determine
$\tilde{\Lambda}_{\overline{MS}}$.

Notice the fundamental significance of the NLO
perturbative coefficient $r_{1}^{\overline{MS}}(Q)$ in all of this.
Knowledge of the
full behaviour of $R(Q)$ at large $Q$ would allow one to extract
$\overline{\Lambda}$ from equation \eq{Lbar}, but this by itself is
{\it observable-dependent} and so one would not test QCD. A test is
only possible if one also knows $r_{1}^{\overline{MS}}(Q)$ (from
Feynman diagram calculations) and can then obtain the {\it universal}
$\tilde{\Lambda}_{\overline{MS}}$ from equation \eq{Lbara}.

The universal $\tilde{\Lambda}_{\overline{MS}}$ of course depends on
the number of active quark flavours,
$\tilde{\Lambda}^{(N_{f})}_{\overline{MS}}$; the RS invariants $b$ and $c$
occurring in equation \eq{FRQ} also involve $N_{f}$. Since
$\tilde{\Lambda}_{\overline{MS}}$ in equation \eq{FRQ} involves the $Q
\rightarrow \infty$ behaviour of $R(Q)$, one might wonder what value for
$N_{f}$ should be taken. With three generations of quarks, $N_{f}=6$
presumably represents the asymptotic number of active flavours and so,
introducing the notation $b(N_{f})$, $c(N_{f})$, the function
$\eta(x)$ in equation \eq{constant} should be chosen as
$\eta(x)=-b(6)x^{2}(1+c(6)x)$.

So finally \eq{FRQ} with the explicit $N_{f}$ dependence exhibited
becomes
\begin{eqnarray}
\label{becomes}
& &\left\{ \frac{1}{R(Q)} + c(6)\ln \left[ \frac{c(6)R(Q)}{1+c(6)R(Q)}
\right] \right\}  \\
& = &b(6)\ln
\frac{Q}{\tilde{\Lambda}^{(6)}_{\overline{MS}}} -
r_{1}^{\overline{MS}}(Q,N_{f}=6) - \int_{0}^{R(Q)} dx \left[
\frac{b(6)}{\xi(x)} + \frac{1}{x^{2}(1+c(6)x)} \right].\nonumber
\end{eqnarray}
Obviously the strict asymptotic $Q \rightarrow \infty$ behaviour of
$R(Q)$ and ${dR}/{d \ln Q} = \xi(R(Q))$ is an idealized concept since
we could never actually measure it. For instance, it presumably makes
no sense to consider QCD in isolation beyond the GUT energy scale.
Rather than using \eq{becomes} for LEP observables where we have
$N_{f}=5$ active flavours, it makes more sense to consider the
equation for $R(Q)$ in a world with $N_{f}=5$  active
flavours, replacing $b(6)$, $c(6)$, and
$\tilde{\Lambda}^{(6)}_{\overline{MS}}$ in \eq{becomes} by $b(5)$, $c(5)$, and
$\tilde{\Lambda}^{(5)}_{\overline{MS}}$. The decoupling theorem
\cite{appelquist}
means that we can consider an $N_{f}=5$ version of QCD for energies
below top threshold. As we shall discuss we shall be interested in how
the observed $Q$-evolution of $R(Q)$ at LEP energies compares with its
asymptotic $Q \rightarrow \infty$ evolution in a {\it hypothetical}
world with $N_{f}=5$ flavours. This will decide how accurately we can
determine the universal constant
$\tilde{\Lambda}^{(5)}_{\overline{MS}}$. To have any chance of
determining  $\tilde{\Lambda}^{(6)}_{\overline{MS}}$ we would require
measurements around the top threshold and at higher energies, which we
do not currently possess.

A further subtlety connected with the effective charge formalism
concerns the fact that one can always define effective charges
other than $a=R$. More generally one could define $f(a)=R$ where
$f(a)= a + f_{2}a^{2} + f_{3}a^{3} + \ldots$ is any analytic function
of $a$. Ch\'{y}la has suggested that this arbitrariness is just the
scheme dependence problem in disguise \cite{jchyla}. In our view this
is incorrect (see also Grunberg's response \cite{grunberg2}). In fact
the arbitrariness in defining the effective charge is completely
equivalent to redefining the observable to be $\tilde{R}=f^{-1}(R)$,
maintaining the standard definition of effective charge $a=\tilde{R}$.
One can of course always consider $\tilde{R}$ as the measured observable
rather than $R$, and then NLO truncation of the perturbation series for
$\tilde{R}$ using some RS will give different results from those
obtained with $R$. Such a redefinition of the observable would only be
useful if some information on the uncalculated higher order
corrections, i.e. knowledge of the function $\rho(R)$, were  available
to inform the choice.

\newpage
{\bf B. EC formalism in NLO - the $\Delta\rho_{0}$ plot.}
\vspace{1cm}

Let us suppose that we have measurements for a number of LEP
observables at a fixed energy ($Q=M_{Z}$), and knowledge of their NLO
QCD perturbative corrections $r_{1}^{\overline{MS}}(Q)$. What can we
learn?

Recall from equation \eq{FRQ} that
\begin{eqnarray}
F(R(Q)) & = & b \ln \frac{Q}{\tilde{\Lambda}^{(5)}_{\overline{MS}}} -
r_{1}^{\overline{MS}}(Q) - \Delta \rho_{0}(Q) \nonumber \\
& = & \rho_{0}(Q) - \Delta \rho_{0}(Q),  \
\end{eqnarray}
where $b$, $c$ and $r_{1}^{\overline{MS}}(Q)$ are evaluated with $N_{f}=5$. As
$Q \rightarrow \infty$, $\Delta\rho_{0}(Q) \rightarrow 0$ and
\begin{equation}
\label{FRQ2}
F(R(Q)) \approx \rho_{0}(Q).
\end{equation}
At sufficiently large $Q$, $\Delta \rho_{0} \ll \rho_{0}$ and \eq{FRQ2} may
be used to obtain $\tilde{\Lambda}^{(5)}_{\overline{MS}}$. We have
that
\begin{equation}
\tilde{\Lambda}^{(5)}_{\overline{MS}} = Q \exp [ -(F(R(Q)) +
r_{1}^{\overline{MS}}(Q) + \Delta \rho_{0}(Q))/b ].
\end{equation}

Given only a NLO perturbative calculation we have no information on
$\Delta \rho_{0}(Q)$. From the measured observable $R_{exp}$ and the
NLO perturbative coefficient we can then extract
\begin{equation}
\label{expd}
\exp (\Delta \rho_{0}(Q)/b) \tilde{\Lambda}^{(5)}_{\overline{MS}} = Q
\exp [-(F(R_{exp}(Q)) + r_{1}^{\overline{MS}}(Q))/b].
\end{equation}
To relate this to the discussions of section 2, the quantity on the
left of equation \eq{expd} is just the value of
$\tilde{\Lambda}_{\overline{MS}}$ which would be extracted from the
data at NLO choosing $\mu = M_{Z} \exp (-r_{1}^{\overline{MS}}/b) =
\mu_{EC}$, the effective charge scale. If $\Delta \rho_{0}(Q) \ll 1$,
then this NLO estimate will be accurate. Indeed the fractional error
in $\tilde{\Lambda}_{\overline{MS}}$ will be
\begin{equation}
\left | \frac{\delta \Lambda}{\Lambda} \right| = \left | 1 - \exp
\left( \frac{\Delta \rho_{0}(Q)}{b} \right) \right | \approx
\frac{\Delta \rho_{0}(Q)}{b}.
\end{equation}
The size of $\Delta \rho_{0}(M_{Z})$ will depend on the QCD
observable. If the $\tilde{\Lambda}^{(5)}_{\overline{MS}}$ values
obtained from equation \eq{expd} exhibit significant scatter, then this
indicates that $\Delta \rho_{0}(Q)$ is not negligible for at least
some of the observables. Recall that $\Delta \rho_{0}(Q)$ is defined
non-perturbatively in
terms of the running of $R(Q)$ by equation (3.8) where $ {dR}/{d \ln
Q} = \xi (R(Q)) = -b\rho(R(Q))$. If $Q$ is sufficiently large that
${dR}/{d \ln Q}$ has its asymptotic ($Q \rightarrow \infty$)
$Q$-dependence, that is if $\rho(R) \approx R^{2}(1+cR)$, then
$\Delta \rho_{0}(Q) \approx 0$. Thus significant scatter in the
$\tilde{\Lambda}^{(5)}_{\overline{MS}}$ extracted at NLO with the
effective charge scale unambiguously indicates that the current
experimental $Q$ is not yet large enough for all observables to be
evolving with $Q$ asymptotically. If we want to reliably determine
$\tilde{\Lambda}^{(5)}_{\overline{MS}}$ additional information on the
`sub-asymptotic' effects is required.

The presence of these effects is a physical fact which can be directly
observed from the $Q$-dependence of data, and cannot be remedied by
changing the unphysical renormalization scheme. We stress once again
how different this is from the scatter obtained with $\mu=M_{Z}$ which
will be partly due to different values of
$r_{1}^{\overline{MS}}(M_{Z})$ and hence which can be modified by changing
the unphysical subtraction procedure.

To make this distinction more precise, let us define
$\tilde{\Lambda}(NLO,r_{1})$ to be the
$\tilde{\Lambda}^{(5)}_{\overline{MS}}$ value extracted at NLO using
an RS with $r_{1}^{\overline{MS}}(\mu)=r_{1}$. The NLO coefficient
$r_{1}$ may be taken to label the RS at NLO. Using $a$ to denote the
coupling constant in this RS we have
\begin{equation}
R=a+r_{1}a^{2},
\end{equation}
and
\begin{equation}
b \ln \frac{\mu}{\tilde{\Lambda}(NLO,r_{1})}=F(a).
\end{equation}
Then using equations (3.20),(3.23), and (3.24) it is straightforward
to show that
\begin{eqnarray}
\tilde{\Lambda}(NLO,r_{1}) & = & \exp [ ( \Delta F(r_{1},R) + \Delta
\rho_{0})/b ]\tilde{\Lambda}^{(5)}_{\overline{MS}} \nonumber \\
& = & \exp(\Delta F(r_{1},R)/b) \tilde{\Lambda}(NLO,0), \
\end{eqnarray}
where
\begin{equation}
\Delta F(r_{1},R) \equiv F(R) - F \left(
\frac{-1+\sqrt{1+4r_{1}R}}{2r_{1}} \right) + r_{1},
\end{equation}
\begin{displaymath}
\Delta F(0,R)=0.
\end{displaymath}
$\tilde{\Lambda}(NLO,0)$ is just the
$\tilde{\Lambda}^{(5)}_{\overline{MS}}$ extracted choosing
$\mu=\mu_{EC}$ ($r_{1}=0$), given by equation (3.21).

 From equation (3.25) we see that the NLO scatter in
$\tilde{\Lambda}^{(5)}_{\overline{MS}}$
is factorized into two components. The first is an {\it already known}
$r_{1}$-dependent contribution involving $\Delta F(r_{1},R)$ which may
be obtained exactly from (3.26) with $R=R_{exp}$, and does {\it not}
involve the unknown NNLO and higher RS invariants. The second component,
$\Delta \rho_{0} \approx \rho_{2}R$,
does not depend on $r_{1}$ but does involve the unknown higher
invariants, and represents the irreducible uncertainty in determining
$\tilde{\Lambda}_{\overline{MS}}$ at NLO. The uninformative
$r_{1}$-dependent part of the scatter can therefore be completely
removed by choosing $\mu=\mu_{EC}$, which sets $\Delta F=0$. The
scatter is then completely given by $\Delta \rho_{0}$.

Rather than focusing on the scatter in
$\tilde{\Lambda}_{\overline{MS}}$, we prefer for presentational reasons
to concentrate on the relative size of $\Delta \rho_{0}$. We can
define for each observable
\begin{equation}
\Delta \rho_{0}^{exp}(Q) \equiv b \ln
\frac{Q}{\tilde{\Lambda}^{(5)}_{\overline{MS}}} -
r_{1}^{\overline{MS}}(Q) - F(R_{exp}(Q)).
\end{equation}
Given a NLO perturbative calculation of $r_{1}^{\overline{MS}}(Q)$ and
experimental measurements of $R_{exp}$ we can then measure $\Delta
\rho_{0}$ from the data up to $b \ln
\tilde{\Lambda}^{(5)}_{\overline{MS}}$ which should be a universal
constant. Changing $\tilde \Lambda_{\overline{MS}}$ merely translates
the $\Delta \rho_{0}$ obtained up or down by the {\it same amount} for
each observable, so we can choose an arbitrary reference value for
$\tilde{\Lambda}^{(5)}_{\overline{MS}}$. For two different observables
$A$ and $B$, the difference $\Delta \rho_{0}{}^{exp}{}_{A} - \Delta
\rho_{0}{}^{exp}{}_{B}$ can then be measured absolutely and so the {\it
relative} size of $\Delta \rho_{0}$ may be investigated. If $\Delta
\rho_{0} \approx 0$ for each observable, corresponding to small
sub-asymptotic effects, then the $\Delta \rho_{0}^{exp}$ points should
lie on a horizontal straight line with some small scatter. Substantial
deviations from a horizontal straight line therefore unambiguously
indicate the presence of sizeable sub-asymptotic effects. The scatter
in $\Delta \rho_{0}^{exp}$ is of course entirely equivalent to the
scatter in the $\tilde{\Lambda}^{(5)}_{\overline{MS}}$ extracted at NLO
with $\mu = \mu_{EC}$, these correspond to the reference
$\tilde{\Lambda}^{(5)}_{\overline{MS}}$ values for which $\Delta
\rho_{0}^{exp}=0$. We shall return to these
$\tilde{\Lambda}^{(5)}_{\overline{MS}}$ values later.

If one has a NNLO perturbative calculation or measurement of $R_{exp}$
at more than one energy then, as we shall discuss in the next
subsection, one can attempt to estimate $\Delta \rho_{0}$ directly.
One can then see if for all observables a single value of
$\tilde{\Lambda}^{(5)}_{\overline{MS}}$ brings $\Delta \rho_{0}^{exp}$
into agreement with the directly estimated $\Delta \rho_{0}$.

We now turn to extracting $\Delta \rho_{0}^{exp}$ from LEP data. The
observables we shall use are the ones appearing in table 1. $R_{2}$
and $R_{3}$ denote the jet fractions defined by
$R_{2}=\sigma_{2}(y_{c})/\sigma_{tot}$ and $R_{3}=
\sigma_{3}(y_{c})/ \sigma_{tot}$, where $\sigma_{2}(y_{c})$ and
$\sigma_{3}(y_{c})$ are the cross-sections for $e^{+}e^{-} \rightarrow
2,3$ resolved jets respectively, with $y_{c}$ the jet resolution cut
and $\sigma_{tot}$ the total hadronic cross-section. The E0 and D
labels denote different recombination algorithms used to cluster
hadrons together. Details of these algorithms and results for the NLO
perturbative coefficients $r_{1}^{\overline{MS}}(Q)$ can be found in
reference \cite{kunst} for the E0 algorithm, and for the D (Durham or
$k_{T}$) algorithm in reference \cite{bks}.

$T$ denotes the thrust variable, and $\chi$ the energy-energy
correlation (EEC) angle. We used recent OPAL data on the differential
distributions in these quantities \cite{opal2,opal4}. The asymmetry in the
energy-energy correlation (AEEC) obtained by subtracting the EEC
measured at $\chi$ and ($180^{o}-\chi$) was also considered. Details
of the definitions of these quantities and their NLO perturbative
coefficients are contained in reference \cite{kunst}.

 From the measured hadronic width of the $Z^{0}$ ($\Gamma_{had}$) and the
$Z^{0}$ leptonic decay width ($\Gamma_{lep}$) one can obtain $R_{Z}
\equiv \Gamma_{had}/ \Gamma_{lep} = (19.97 \pm 0.03)(1 +
\delta_{QCD})$, where $\delta_{QCD}$ has a perturbation series of the
form (2.1) with $r_{1}^{\overline{MS}}(M_{Z})=1.41$. The electroweak
contribution ($19.97 \pm 0.03$) is from reference \cite{bardin}. This
coefficient is for massless quarks, if one includes heavy quark
masses \cite{chetyrkin}, one finds $\delta_{QCD}=(1.05a + 0.9 a^{2})$ instead
in the $\overline{MS}$ scheme with $\mu=M_{Z}$. Within the
errors the values of $\Delta \rho_{0}^{exp}$ obtained are
are consistent, so we shall use massless QCD. For the
experimental value we take the 1993 LEP average $R_{Z}= 20.763 \pm
0.049$ \cite{newrz}.

As noted earlier the reference value of $\tilde
\Lambda^{(5)}_{\overline{MS}}$ assumed for $\Delta \rho_{0}^{exp}$ is
irrelevant since we are concentrating here on the scatter of $\Delta
\rho_{0}^{exp}$ values.  We shall choose
$\tilde\Lambda^{(5)}_{\overline{MS}}=110$ MeV which corresponds to the
central value obtained from a non-perturbative lattice analysis of
the 1P-1S splitting in the charmonium system \cite{el}.
Whilst at present limited by the use of a quenched approximation, such
calculations could in future provide reliable estimates of $\tilde \Lambda$.
In Figure 3 we
show the $\Delta \rho_{0}^{exp}$ values obtained for the LEP
observables discussed above. We have taken for each observable the
particular values of the kinematical variables given in Table 1. Of
course for each observable we actually have a $\Delta \rho_{0}^{exp}$
distribution in the associated kinematical variable. In Figures 4-7 we
give the $\Delta \rho_{0}^{exp}$ distributions for the jet rates
$R_{2}(E_{0})$, $R_{3}(E_{0})$, $R_{2}(D)$, $R_{3}(D)$ as functions of
the jet resolution cut $y_{c}$. We see that for $y_{c}
\gsim 0.04$, $\Delta \rho_{0}$ is constant within the errors. For
smaller $y_{c}$ there is a much stronger dependence. In this small $y_{c}$
region we may expect sizeable corrections involving $\ln y_{cut}$. An
attempt to resum leading and next-to-leading logarithms to all-orders
for $\Delta \rho_{0}$ for $R_{2}(D)$ will be discussed in sub-section
3D. In Figures 4-7 we used
published OPAL data \cite{opal2} without correcting for hadronization effects.
These
are expected to be totally negligible for $y_{c} \gsim 0.04$, but
will be important in the small $y_{c}$ region. For thrust and EEC
the $\Delta \rho_{0}^{exp}$ distributions are flat away from $T \approx 1$
and $\chi \approx 180^{o}$ where large corrections involving $ \ln(1-T)$
and $\ln(\cos^{2} \frac{\chi}{2})$ respectively will be important.

Returning to Figure 3 we see that there is significant scatter, in
particular between the jet rates with E0 and D recombination algorithms.
To proceed further then we need either NNLO calculations or information
on the $Q$-dependence of these observables. The only NNLO calculation
completed so far is that for $R_{Z}$ \cite{gorishny}.
For $R_{3}(E0)$ and EEC there are published data from the JADE collaboration at
the PETRA $e^{+}e^{-}$ machine with $\sqrt{s}$ in the range 22-44 GeV
\cite{jade,jade2}, using a comparable analysis. In the next sub-section we
shall
attempt to use this data together with the LEP data for these quantities
to obtain absolute estimates of $\Delta \rho_{0}$ from the $Q$-dependence.

We conclude by tabulating in Table 2 the
$\tilde{\Lambda}^{(5)}_{\overline{MS}}$ values obtained by fitting to LEP
data at NLO for these same observables at the same values of the kinematical
parameters as in Table 1 using $\mu = M_{Z}$ and $\mu = \mu_{EC}$. The
$\mu = \mu_{EC}$ values correspond to the
$\tilde{\Lambda}^{(5)}_{\overline{MS}}$ values for which $\Delta
\rho_{0}=0$, and so the scatter is directly related to points in the
$\Delta \rho_{0}$
plot of Figure 3. The $\mu=M_{Z}$ values show much greater scatter, since
as discussed in section 2 the scatter in $r_{1}^{\overline{MS}}(M_{Z})$ is
superimposed on top of the physically-relevant scatter in $\Delta \rho_{0}$.

To further emphasise this, suppose that we lived in an `asymptotic world'
where $Q=M_{Z}$ was sufficiently large that $\Delta \rho_{0} \approx 0$ for all
observables. Thus for all observables
\begin{equation}
\label{FRexp}
F(R_{exp}(M_{Z})) = b \ln \frac{M_{Z}}{\tilde{\Lambda}^{(5)}_{\overline{MS}}}
- r_{1}^{\overline{MS}}.
\end{equation}
Supposing that $\tilde{\Lambda}^{(5)}_{\overline{MS}}=200$ MeV, then \eq{FRexp}
may be used to generate data `$R_{exp}(M_{Z})$' in such an
`asymptotic world'. By construction the
$\tilde{\Lambda}^{(5)}_{\overline{MS}}$ obtained by fitting at NLO with
$\mu=\mu_{EC}$ will be 200 MeV for all observables, but that obtained with
$\mu=M_{Z}$ will still exhibit significant, but completely
predictable,scatter. These `asymptotic
world' results are also tabulated in Table 2.

\vspace{1cm}
{\bf C. $\Delta \rho_{0}$ from NNLO calculations and Q-dependence.}
\vspace{1cm}

If we have available a NNLO perturbative calculation then the RS invariant
$\rho_{2}$ defined in equation (2.28) can be obtained. For $R_{Z}$,
$\rho_{2}=-15.1$ ($N_{f}=5$) using the NNLO calculation of reference
\cite{gorishny}. For all the other observables NNLO corrections have yet to be
calculated and so $\rho_{2}$ is unknown. If we insert the perturbative
expansion of $\rho(x)$ (equation (2.26)) into (3.8) we obtain
\begin{eqnarray}
\Delta \rho_{0} & = & \int^{R}_{0} dx \frac{(\rho_{2}+\rho_{3}x+\ldots)}
{(1+cx)(1+cx+\rho_{2}x^{2}+\ldots)} \nonumber \\
& \approx & \rho_{2}R + ( \frac{\rho_{3}}{2}-c \rho_{2})R^{2} +
O(R^{3})
\equiv \Delta \rho_{0}^{NNLO}+O(R^{2}). \
\end{eqnarray}
Using the 1993 LEP average data we have $R_{Z} = (19.97 \pm 0.03)
(1+\delta_{QCD})$ with $\delta_{QCD}=0.040 \pm 0.004$ \cite{newrz}. Thus with
$\rho_{2}=-15.1$ we have $\Delta \rho_{0}^{NNLO} \simeq \rho_{2} \delta_{QCD}
\simeq -0.60 \pm 0.06$. Adjusting the reference value of
$\tilde{\Lambda}^{(5)}_{\overline{MS}}$ so that $\Delta \rho_{0}^{exp}$
for $R_{Z}$ corresponds with this $\Delta \rho_{0}^{NNLO}$ results in
$\tilde{\Lambda}^{(5)}_{\overline{MS}}=288 \pm 200$ MeV. This is of
course just the $\tilde{\Lambda}_{\overline{MS}}$ obtained at NNLO
using the EC scheme with $\tau=\rho_{0}$ and $c_{2}=\rho_{2}$.

To obtain estimates of $\Delta \rho_{0}$ for the other observables one
can try to use their $Q$-dependence. Suppose that we have measurements of $R$
at
two energies $Q=Q_{1}$ and $Q=Q_{2}$ ($Q_{1} > Q_{2}$), then we can construct
$\Delta \rho_{0}^{exp}(Q_{1})$ and $\Delta \rho_{0}^{exp}(Q_{2})$ of equation
(3.27) from the data (with the same reference value of
$\tilde{\Lambda}^{(5)}_{\overline{MS}}$). Then
\begin{eqnarray}
\Delta \rho_{0}^{exp}(Q_{1}) - \Delta \rho_{0}^{exp}(Q_{2}) & = &
b \ln \frac{Q_{1}}{Q_{2}} - F(R(Q_{1})) + F(R(Q_{2})) \nonumber \\
& = & \int^{R(Q_{1})}_{R(Q_{2})} I(x) dx = (R(Q_{1})-R(Q_{2}))I(\overline{R}),
 \
\end{eqnarray}
where
\begin{equation}
I(x) \equiv \left[ -\frac{1}{\rho(x)} + \frac{1}{x^{2}(1+cx)} \right]
\end{equation}
is the integrand of $\Delta \rho_{0}$ and $R(Q_{1})<\overline{R}<R(Q_{2})$.
Thus the integrand of $\Delta \rho_{0}$ may be measured from the $Q$-dependence
of the data,
\begin{equation}
I(\overline{R}) = \frac{\Delta \rho_{0}^{exp}(Q_{1}) - \Delta \rho_{0}^{exp}
(Q_{2})}{(R(Q_{1})-R(Q_{2}))}.
\end{equation}
Obviously by measuring ${dR}/{d \ln Q}$ at $Q=Q_{0}$ with sufficient
accuracy one can in principle determine $I(R(Q_{0}))$ and so by suitably
detailed measurement over the energy range $[Q_{1},Q_{2}]$ one can determine
$I(x)$ on the interval $[R(Q_{1}),R(Q_{2})]$.

The situation for measurements at $Q_{1}$ and $Q_{2}$ is shown in Figure 8
for $I(x)$ versus $x$. The uncertainty in $\overline{R}$ is represented by the
horizontal error bar between $R(Q_{1})$ and $R(Q_{2})$ and the measurement
errors in $R(Q_{1})$ and $R(Q_{2})$ themselves contribute a vertical error
bar.

We have from equation (3.29) that
\begin{equation}
I(x)=\rho_{2} + (\rho_{3}-2c \rho_{2})x + O(x^{2}).
\end{equation}
So $I(0)=\rho_{2}$. Thus from a NNLO perturbative calculation we can obtain
the integrand at the origin. Notice that, like $\overline{\Lambda}$, the RS
invariant $\rho_{2}$ is connected with the asymptotic $Q$-dependence of $R(Q)$.
A next-NNLO calculation would provide $\rho_{3}$ and tell us the slope
of $I(x)$
at the origin, $\rho_{3}-2c\rho_{2}$.

If NNLO perturbation theory is adequate to determine $\Delta \rho_{0}$, then
$\Delta \rho_{0} \approx \Delta \rho_{0}^{NNLO}=\rho_{2}R(Q)$, corresponding
to $I(x) \approx \rho_{2} =$ constant, on the range $[0,R(Q)]$. Thus,
if
NNLO perturbation theory is adequate, then we expect $I(\overline{R}) \approx
\rho_{2}$ and so from the measurements of $R(Q_{1})$ and $R(Q_{2})$ we can
estimate $\Delta \rho_{0}^{est}(Q_{1}) = I(\overline{R})R(Q_{1})$ from (3.32),
\begin{equation}
\Delta \rho_{0}^{est}(Q_{1}) = \frac{(\Delta \rho_{0}^{exp}(Q_{1}) -
\Delta \rho_{0}^{exp}(Q_{2}))}{(R(Q_{1})-R(Q_{2}))}R(Q_{1}).
\end{equation}

If NNLO calculations eventually become available for the observables one can
then check explicitly whether $\Delta \rho_{0}^{NNLO} \approx \Delta
\rho_{0}^{est}$.
A marked discrepancy would indicate the importance of next-NNLO and
higher perturbative effects and/or large `non-perturbative' $e^{-1/R}$ effects.
Given $\rho_{2}$ one can then estimate $\rho_{3}$ from the slope of the
straight line joining $I(0)$ and $I(\overline{R})$ and obtain an improved
estimate of $\Delta \rho_{0}^{est}(Q_{1})$ from the area under the
trapezium (see Figure 8).

In this way perturbative calculations and experimental $Q$-dependence
measurements serve as complementary pieces of information. The NNLO and higher
perturbative calculations effectively provide details of the
$Q \rightarrow \infty$ running of $R(Q)$ which could never be obtained
experimentally.
Together they can help to constrain the behaviour of the function
$\xi(R)=-b\rho(R)$,
and hence to refine the estimates of $\tilde{\Lambda}^{(5)}_{\overline{MS}}$.

The running of the observables with energy has been used before as a test of
QCD. For instance $R_{3}(E0,y_{c}=0.08)$ has been studied as a function
of $Q$ over the PETRA-LEP energy range \cite{bethke}. The apparent running has
been
compared with the NLO QCD expectations for various choices of renormalization
scale $\mu$, and $\tilde{\Lambda}$. It has even been suggested \cite{jezabek}
that
the LEP measurements may indicate slightly less running than is expected from
QCD, and that light gluinos, which would modify the QCD $\beta$-function,
cannot be excluded. Such a claim seems ludicrously premature given
our lack of knowledge of NNLO and higher QCD effects.

If we take data for $R_{3}(E0,y_{c}=0.08)$ at $Q_{2}=34$ GeV from JADE
\cite{jade},
and at $Q_{1}=91$ GeV from OPAL \cite{opal2}, then we obtain from (3.34),
$\Delta \rho_{0}^{est}(91)=-5.5 \pm 3$, which corresponds to $I(\overline{R}) =
-110 \pm 60$. So if NNLO perturbation theory is adequate for this observable
we estimate $\rho_{2}=-110 \pm 60$, for the as yet uncalculated NNLO RS
invariant.
For the EEC with $\chi=60^{o}$ measured by JADE at $Q_{2}=34$ GeV \cite{jade2},
and
by OPAL at $Q_{1}=91$ GeV \cite{opal4} we similarly find $\Delta
\rho_{0}^{est}(91)=-1.5 \pm 1.5$,
which corresponds to $I(\overline{R})=-21 \pm 21$, and an estimate of
$\rho_{2}=-21 \pm 21$. Since $\Delta \rho_{0}^{exp}$ is very insensitive to
$y_{c}$ and $\chi$, comparable results are obtained for other choices of these
variables.

In figure 9 we plot $\Delta \rho_{0}$ for $R_{Z}$,
$R_{3}(E0,y_{c}=0.08)$ and
$R_{EEC}(\chi=60^{o})$ with $Q=M_{Z}$.  The
diamonds correspond to absolute predictions. For $R_{Z}$ we take
$\Delta \rho_{0}^{NNLO}=-0.60 \pm 0.06$, and $R_{3}(E0,y_{c}=0.08)$
and for
$R_{EEC}(\chi=60^{0})$ the $\Delta \rho_{0}^{est}(91)$ values noted above
obtained from $Q$-dependence. The crosses correspond to the $\Delta
\rho_{0}^{exp}$ for these observables with the reference
$\tilde{\Lambda}^{(5)}_{\overline{MS}}$ chosen as 287 MeV so that
$\Delta \rho_{0}^{exp}$ for $R_{Z}$ agrees with $\Delta \rho_{0}^{NNLO}$.
There is then agreement, within the considerable errors, between
$\Delta \rho_{0}^{exp}$ and $\Delta \rho_{0}^{est}$ for $R_{3}$ and $R_{EEC}$.
Demanding consistency between $\Delta \rho_{0}^{exp}$ and $\Delta
\rho_{0}^{est}$,
$\Delta \rho_{0}^{NNLO}$ then requires $\tilde{\Lambda}^{(5)}_{\overline{MS}}
=287 \pm 100$ MeV.

Although only performed for a limited number of observables, we regard this
consistency between $\Delta \rho_{0}^{exp}$ and the absolute $\Delta \rho_{0}$
obtained from NNLO calculations and $Q$-dependence measurements as very
encouraging. Ideally one would like to see additional LEP measurements for
all observables off the $Z$ peak at a lower value of $Q$ to avoid the
uncertainties due to combining measurements from different machines and
detectors. One could then perform an exhaustive analysis of the kind described
here and obtain a reliable determination of
$\tilde{\Lambda}^{(5)}_{\overline{MS}}$. Scatter in the
$\tilde{\Lambda}^{(5)}_{\overline{MS}}$ values obtained by adjusting
$\Delta \rho_{0}^{exp}$ to agree with $\Delta \rho_{0}^{est}$ (obtained from
$Q$-dependence using (3.34)) would then unambiguously indicate the importance
of next-NNLO corrections or non-perturbative $e^{-1/R}$ effects for some of
the observables.

It is interesting that the above $Q$-dependence estimates of $\rho_{2}$, and
the exact NNLO calculation of $\rho_{2}$ for $R_{Z}$, suggest $\rho_{2}$
large ($O(10)$) and {\it negative}. This has  implications for the
infra-red behaviour of $R(Q)$ since it is consistent with fixed-point
behaviour, that is $\rho(R^{*})=0$ implying $R \rightarrow R^{*}$ as
$Q \rightarrow 0$ \cite{chyla,matt}.

\vspace{1cm}
{\bf D. $\Delta \rho_{0}$ from next-to-leading logarithm resummations}
\vspace{0.5cm}

There has been much recent interest, both theoretical
\cite{catani,catani2} and
experimental \cite{opal1}, in the possibility of resumming leading and
next-to-leading logarithms (NLL) in kinematical variables to all-orders for
LEP observables. By identifying the leading and next-to-leading logarithms in
the perturbative coefficients $r_{k}=A_{k}L^{2k}+B_{k}L^{2k-1}+\ldots$, with
$L \equiv \ln \frac{1}{\lambda}$ (where $\lambda$ is $y_{c},1-T,
\cos^{2}\frac{\chi}{2},\ldots$), and by demonstrating generalised
exponentiation one can resum these terms to all-orders. To include the exact
NLO perturbative coefficient, however, some ad hoc matching prescription is
then required \cite{opal1,catani2}, and anyway the problem of renormalization
scale dependence
still remains.

The advantage of the EC formalism here is that from the leading and
next-to-leading
logarithms in $r_{k}$ one can obtain the leading and next-to-leading
logarithms in the RS invariants $\rho_{k}=\tilde{A}_{k}L^{2k}+
\tilde{B}_{k}L^{2k-1}+O(L^{2k-2})$. One can then {\it unambiguously} construct
\begin{eqnarray}
\rho^{NLL}(x) &=& x^{2}(1+cx+\rho_{2}^{NLL}x^{2}+\ldots+\rho_{k}^{NLL}+\ldots),
\nonumber \\
\rho_{k}^{NLL} & \equiv & \tilde{A}_{k}L^{2k}+ \tilde{B}_{k}L^{2k-1}. \
\end{eqnarray}
$\rho^{NLL}(x)$ is constructed from RS-invariants and so does not involve the
renormalization scale, or the NLO perturbative coefficient. There is,
therefore, no analogue of the matching prescription ambiguity and one can
obtain
\begin{equation}
\Delta \rho_{0}^{NLL} \equiv \int^{R_{exp}(Q)}_{0} dx \left[ -
\frac{1}{\rho^{NLL}(x)} + \frac{1}{x^{2}(1+cx)} \right].
\end{equation}
This may be directly compared with $\Delta \rho_{0}^{exp}$. If the observed
$\lambda$-dependence of $\Delta \rho_{0}^{exp}$ in the small-$\lambda$
region disagrees with that predicted from $\Delta \rho_{0}^{NLL}$ then one
has unambiguous evidence that the NLL approximation is inadequate, that is
that the neglected sub-leading logarithms and constants are important.

We shall perform this comparison here only for the 2-jet rate $R_{2}(D)$,
where the exponentiation of logarithms is rather straightforward, deferring
discussions of other observables. We have
\begin{equation}
R_{2}(D)=1+r_{0}R
\end{equation}
where $R$ has a perturbation series of the form (2.1). To NLL accuracy
\cite{catani},
\begin{equation}
R_{2}(D) = \exp \left[ \frac{3C_{F}L}{2} \left( 1-\frac{L}{3} \right)a
- b \frac{C_{F}}{6}L^{3}a^{2} \right]
\end{equation}
where $a=a(\mu=Q),L=\ln \frac{1}{y_{c}}$. The $b$ dependent term in the
exponent may be absorbed into $a$ by a change of scale $Q \rightarrow
y_{c}^{1/3}Q$. Since the RS invariants $\rho_{k}$ are independent of the
renormalization scale to construct them, we can equally use
\begin{equation}
R_{2}(D)=\exp \left[ \frac{3C_{F}L}{2} \left( 1-\frac{L}{3} \right)a \right]
\equiv \exp(r_{0}^{NLL}a).
\end{equation}
One has from (3.37) that
\begin{equation}
R=\frac{R_{2}(D)-1}{r_{0}}
\end{equation}
and from (2.27) with $\beta(a)=a^{2}(1+ca)$,
\begin{equation}
\rho(R)=a^{2}(R)(1+ca(R)) \frac{dR}{da}.
\end{equation}
 From the form of $\rho_{k}$ of (2.28) one sees that $c_{2},c_{3},\ldots$ will
not contribute at NLL level. We retain $c$ in $\beta(a)$ so that $\rho^{NLL}(x)
=x^{2}(1+cx+\ldots)$. From (3.39) and (3.40) one has
\begin{equation}
\frac{dR}{da}=\exp(r_{0}^{NLL}a)
\end{equation}
then using (3.41) one finds
\begin{equation}
\rho^{NLL}(R) = a^{2}(R)(1+ca(R)) \exp(r_{0}^{NLL}a(R))
\end{equation}
where
\begin{equation}
a(R) = \frac{\ln (1+r_{0}^{NLL}R)}{r_{0}^{NLL}}.
\end{equation}
If one then inserts $\rho^{NLL}(x)$ of (3.43) into (3.36) and change variables
from $x$ to
\begin{equation}
u(x) \equiv \frac{\ln(1+r_{0}^{NLL}x)}{r_{0}^{NLL}}.
\end{equation}
Performing the integration one finds
\begin{equation}
\Delta \rho_{0}^{NLL} = F(u(R_{exp}))-F(R_{exp}) - \frac{r_{0}^{NLL}}{2}.
\end{equation}

In Figure 10(a) the diamond points show the absolute prediction for
$\Delta \rho_{0}^{NLL}$ versus $y_{cut}$ obtained using (3.46)
together with the OPAL data of reference \cite{opal2} for $R_{2}(D)$
uncorrected for hadronization effects. The other points show the
corresponding $\Delta \rho_{0}^{exp}$ (see Figure 5) with the
reference value of $\tilde{\Lambda}^{(5)}_{\overline{MS}}$ adjusted so
that $\Delta \rho_{0}^{exp}$ agrees with $\Delta \rho_{0}^{NLL}$ at
the peak in $\Delta \rho_{0}^{NLL}$. Notice that for $y_{c} \lsim 0.01$ the NLL
prediction
has a completely different low-$y_{c}$ behaviour from the uncorrected
data. In Figure 10(b) the same quantites using hadronization-corrected
OPAL data for $R_{2}(D)$ \cite{wethank} are displayed. Evidently, the low
$y_{c}$ behaviour is now at least in qualitative agreement. We
conclude that in order to test the adequacy of the NLL approximation
we are having to compare with data in a region where hadronization
corrections are sizeable. Unless one is extremely confident in one's
ability to estimate these corrections it is hard to draw firm conclusions.

%%%%Section 4
\newpage
%%\vspace{0.5cm}
\section{Summary and Conclusions}
%%\vspace{0.5cm}
\indent

 Evidently if one has a NLO perturbative QCD calculation available
there will be some uncertainty in the NLO perturbative prediction for
a QCD observable due to the missing uncalculated higher-order terms
in the perturbation series. This has the effect that when one compares
the NLO calculation with experimental data one may anticipate that the
value of $\tilde{\Lambda}_{\overline{MS}}$ extracted will not be
universal but will exhibit some scatter for different observables due
to the differing sizes of the uncalculated contributions. The problem
is compounded by the fact that the $\tilde{\Lambda}_{\overline{MS}}$
extracted depends also on the choice of RS at NLO, which may be
labelled by $r_{1}$ the NLO coefficient (or equivalently the
renormalisation scale $\mu$), an unphysical parameter. We have shown in
equation (3.25) that the relation between the extracted
$\tilde{\Lambda}(NLO,r_{1})$ and the actual
$\tilde{\Lambda}_{\overline{MS}}$ which one is trying to measure is
factorized into two contributions. One, $\Delta F(r_{1},R)$, is
$r_{1}$-dependent and known exactly from equation (3.26), and the
other, $\Delta \rho_{0}$, is unknown but $r_{1}$-independent. $\Delta
\rho_{0}$ depends only on the observable and NNLO and higher
RS-invariants $\rho_{2},\rho_{3},\ldots$. The predictable
$r_{1}$-dependence of $\tilde{\Lambda}(NLO,r_{1})$ therefore has
nothing to tell us about the importance of uncalculated corrections,
the irreducible uncertainty resides in the unknown $\Delta \rho_{0}$. Some of
the
$r_{1}$-dependent logarithms in the NNLO and higher corrections can be
summed up into $\Delta F(r_{1},R)$ and the remainder can be absorbed
into RS-invariant combinations contained in $\Delta \rho_{0}$.

The unknown $\Delta \rho_{0}$ can be isolated by choosing the
particular RS where $r_{1}=0$, $\mu=\mu_{EC}$ the effective charge
scale, which sets $\Delta F=0$. From the scatter in
$\tilde{\Lambda}(NLO,0)$ for different observables one can then infer
the scatter in $\Delta \rho_{0}$, in particular relative differences
$\Delta \rho_{0A}-\Delta \rho_{0B}$ for observables A,B can be
absolutely measured. Figure 3 indicates that these relative
differences cannot be neglected for a range of LEP observables. For at
least some of the observables $\Delta \rho_{0}$ must be sizeable, and
so $\Delta \rho_{0}$ must be estimated before we can determine
$\tilde{\Lambda}_{\overline{MS}}$ with any reliability.

In contrast, in the standard LEP determinations of $\alpha_{s}(M_{Z})$
one tries to estimate the importance of uncalculated higher-order
corrections by using two different ad hoc scale choices and
interpreting the spread in extracted $\alpha_{s}(M_{Z})$ as indicating
a `theoretical error'. By artificially enlarging uncertainties in this
way one obtains a spurious consistency between different observables,
the real scale-independent uncertainty due to $\Delta \rho_{0}$ being
buried beneath the supposedly informative scale dependence of $\Delta
F$. The global $\alpha_{s}(M_{Z})$ determinations obtained with these
sort of analyses should therefore be treated with some scepticism.

Having determined that $\Delta \rho_{0}$ is not negligible one must
try to estimate it to make further progress. Within the effective
charge formalism one can write $\tilde{\Lambda}_{\overline{MS}}$ in
terms of the measured observable $R(Q)$, $r_{1}^{\overline{MS}}$ and
$\Delta \rho_{0}$ constructed non-perturbatively from the measured
running ${dR}/{d \ln Q}=-b\rho(R)$. $\Delta \rho_{0}$ may also be
obtained perturbatively by expanding the effective charge
$\beta$-function, so that given a NNLO calculation one can estimate
$\Delta \rho_{0}^{NNLO}=\rho_{2}R$. By combining NNLO calculations and
$Q$-dependence measurements one can refine ones knowledge of $\Delta
\rho_{0}$ and hence of $\tilde{\Lambda}_{\overline{MS}}$. The
perturbative RS-invariants
$\rho_{0}(\overline{\Lambda}),\rho_{2},\rho_{3},\ldots$ are connected
with $Q \rightarrow \infty$ evolution of the observable $R(Q)$, and
hence provide information about the function $\rho(R)$ in the vicinity
of $R=0$.

The present situation is that a NNLO calculation is available only for
the hadronic width of the $Z^{0}$. By estimating $\Delta
\rho_{0}^{NNLO}$ for this observable, and combining PETRA data for jet
rates and energy-energy correlations, with LEP data on these
observables, to obtain an estimate of $\Delta \rho_{0}$ from
$Q$-dependence, we found consistency for
$\tilde{\Lambda}^{(5)}_{\overline{MS}}=287 \pm 100$ MeV.

We also showed that for observables where leading and next-to-leading
logarithms in kinematical variables can be resummed one can
unambiguously estimate $\Delta \rho_{0}^{NLL}$ by resumming leading
and next-to-leading logarithms in the RS invariants $\rho_{k}$. In the
conventional approach one needs to use an ad hoc matching procedure to
include the exact NLO perturbative coefficient, and the problem of
scale dependence still remains. Performing this analysis for
$R_{2}(D)$ we found qualitative agreement between the
$y_{c}$-dependence at low $y_{c}$ of $\Delta \rho_{0}^{NLL}$ and
$\Delta \rho_{0}^{exp}$ using hadronization corrected data, but a
completely opposite low $y_{c}$ dependence using uncorrected data.

We conclude that reliable measurements of
$\tilde{\Lambda}_{\overline{MS}}$ at $e^{+}e^{-}$ machines will
require at least NNLO perturbative calculations and/or measurements of
observables at more than one energy. Acquiring such information will
entail considerable experimental and theoretical effort. The effective
charge formalism allows one to efficiently harness this hard won
information to refine one's knowledge of
$\tilde{\Lambda}_{\overline{MS}}$ and the interplay between
perturbative and non-perturbative effects. The insistance on
formulating everything in terms of physical quantities allows one to
quantify uncertainties in a way which is impossible in approaches
which choose the unphysical RS-dependence parameters according to some
plausiblity argument.

An important remaining problem is to extend the approach to processes
with initial state hadrons where there is an additional factorization
ambiguity connected with the separation of structure functions from
the hard cross-sections.

%\newpage
\section*{Acknowledgements}
%%\vspace{1cm}
We would like to thank Siggi Bethke, Nigel Glover, Andrei Kataev, and
Paul Stevenson for stimulating discussions and useful comments which
have helped us to clarify our ideas. Paolo Nason is thanked for
supplying a computer code for generating NLO corrections to the
$e^{+}e^{-}$ matrix elements.

M.T.R. gratefully acknowledges receipt of a S.E.R.C U.K. Studentship.
\newpage
%%\section*{References}

\section*{Figure Captions}
\vspace{0.5cm}

\begin{description}

\item[Figure 1.] The generic approximately inverted-parabolic behaviour
of the NLO approximant $R^{1}(\tau)$ versus $a(\tau)$. The figure is
plotted for $\rho_{0}=25$ and $N_{f}=5$. The dashed line represents
measured data (see text).

\item[Figure 2(a).] $\alpha_{s}(M_{Z})$ obtained by fitting NLO
perturbative calculations with $\mu=M_{Z}$ to OPAL data.

\item[Figure 2(b).] As (a) but with enlarged `theoretical errors' using
$\mu=M_{Z}$ and fits to data, as described in the text. Reproduced
from reference [8].

\item[Figure 3.] $\Delta \rho_{0}^{exp}$ of equation (3.27) for various
LEP observables (see text for details). A reference value of
$\tilde{\Lambda}^{(5)}_{\overline{MS}}=110$ MeV is chosen.

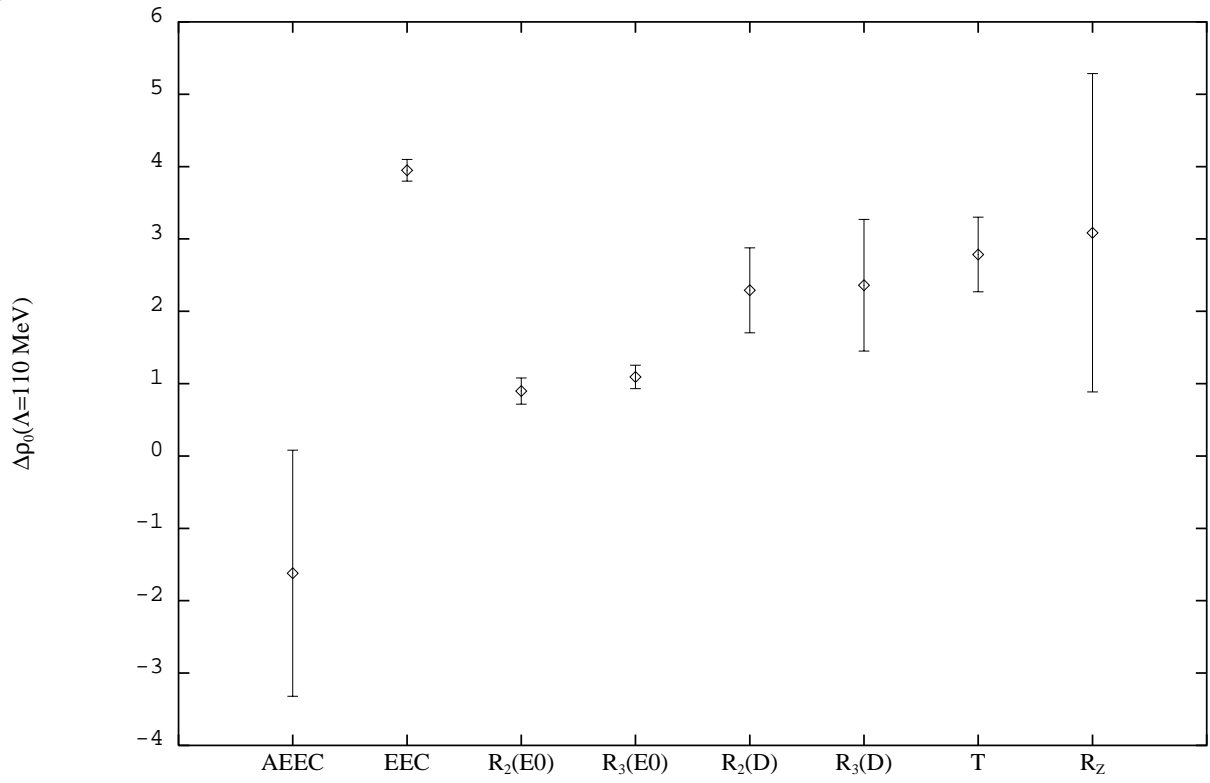
\item[Figure 4.] The $y_{cut}$ dependence of $\Delta \rho_{0}^{exp}$
for $R_{2}(E0)$. OPAL data of reference [2] is used uncorrected for
hadronization. A reference value of
$\tilde{\Lambda}^{(5)}_{\overline{MS}}=110$ MeV is chosen.

\item[Figure 5.] As Figure 4 but for $R_{2}(D)$.

\item[Figure 6.] As Figure 4 but for $R_{3}(E0)$.

\item[Figure 7.] As Figure 4 but for $R_{3}(D)$.

\item[Figure 8.] Learning about the integrand of $\Delta \rho_{0}$,
$I(x)$ of equation (3.31). NNLO perturbative calculations can provide
$I(0)=\rho_{2}$, and measurements at energies $Q_{1}$ and $Q_{2}$ can
provide a point away from the origin.

\item[Figure 9.] Absolute estimates of $\Delta \rho_{0}$ for LEP
observables at $Q=91$ GeV are compared with the observed $\Delta
\rho_{0}^{exp}$ with $\tilde{\Lambda}^{(5)}_{\overline{MS}}=288$ MeV.
For EEC($\chi=60^o$) and $R_{3}(E0,y_{c}=0.08)$ the absolute estimates
are $\Delta \rho_{0}^{est}$ of equation (3.34) obtained from
$Q$-dependence using JADE data. For $R_{Z}$ $\Delta \rho_{0}^{NNLO}$
from the NNLO calculation of reference [12] is used.

\item[Figure 10(a).] For the observable $R_{2}(D)$ the NLL
approximation $\Delta \rho_{0}^{NLL}$ of equation (3.46) versus
$y_{c}$ is compared with $\Delta \rho_{0}^{exp}$ with
$\tilde{\Lambda}^{(5)}_{\overline{MS}}$ adjusted. OPAL data of refervence
[2] uncorrected for hadronization effects is used.

\item[Figure 10(b).] As (a) but with hadronization-corrected data.

\end{description}\newpage
\begin{center}
\begin{tabular}{|c|c|c|}\hline
Observable & &$r_{1}^{\overline{MS}}(M_{Z})$ \\
\hline
T & T=0.9 & 13.41 \\
$R_{2}(E_{0})$ & $y_{cut}=0.08$ & 10.22 \\
$R_{3}(E_{0})$ & $y_{cut}=0.08$ & 10.09 \\
$R_{2}(D)$ & $y_{cut}=0.08$ & 7.23 \\
$R_{3}(D)$ & $y_{cut}=0.08$ & 7.15 \\
EEC & $\chi=60^{o}$ & 11.30 \\
AEEC & $\chi=60^{o}$ & 5.13 \\
$R_{Z}$& & 1.411 \\
%$R_{e^{+}e^{-}}$ & & 1.411 \\
\hline
\end{tabular}
\end{center}
\vspace{0.5cm}
{\bf Table 1:} NLO coefficients $r_{1}^{\overline{MS}}(M_{Z})$ from
[11] for various LEP observables at the particular values of
the kinematical variables noted.
\vspace{1cm}
\begin{center}
\begin{tabular}{|c|c|c|c|c|}\hline
Observable & LEP & LEP & $\Delta \rho_{0} \approx 0$ & $\Delta
\rho_{0} \approx 0 $  \\
& $\Lambda^{(5)}_{\overline{MS}}$ &
$\Lambda^{(5)}_{\overline{MS}}$ & $\Lambda^{(5)}_{\overline{MS}}=200~MeV$ &
$\Lambda^{(5)}_{\overline{MS}}=200~MeV$ \\
& $\mu=M_{Z}$ & $\mu=\mu_{EC}$ & $\mu=M_{Z}$ & $\mu=\mu_{EC}$ \\
\hline
T & 883.0 & 197.9 & 894.4 & 200 \\
$R_{2}(E_{0})$ & 335.7& 140.4 & 498.5 & 200 \\
$R_{3}(E_{0})$ & 326.8 & 139.5 & 488.3 & 200 \\
$R_{2}(D)$ & 420.3 & 252.1 & 328.3 & 200  \\
$R_{3}(D)$ & 425.6 & 257.4 & 325.2 & 200 \\
EEC & 839.1 & 268.2 & 597.4 & 200\\
AEEC & 89.23 & 63.79 & 294.2 & 200 \\
$R_{Z}$& 254.4 & 245.7 & 206.9 & 200 \\
%$R_{e^{+}e^{-}}$ & 1533.7 & 1463.2 & 138.41 & 200 \\
\hline
\end{tabular}
\end{center}
\vspace{0.5cm}
{\bf Table 2:} The first two columns show the central values of
$\tilde{\Lambda}^{(5)}_{\overline{MS}}$ (in MeV) extracted by
comparing NLO perturbative QCD predictions for the observables of
Table 1, taking $\mu=M_{Z}$ and $\mu=\mu_{EC}$, with OPAL data [2].
The third and fourth columns give the NLO
$\tilde{\Lambda}^{(5)}_{\overline{MS}}$ values extracted with
$\mu=M_{Z}$ and $\mu=\mu_{EC}$, assuming $\Delta \rho_{0}=0$ for each
observable, and an actual value of $\Lambda^{(5)}_{\overline{MS}}=200$ MeV.
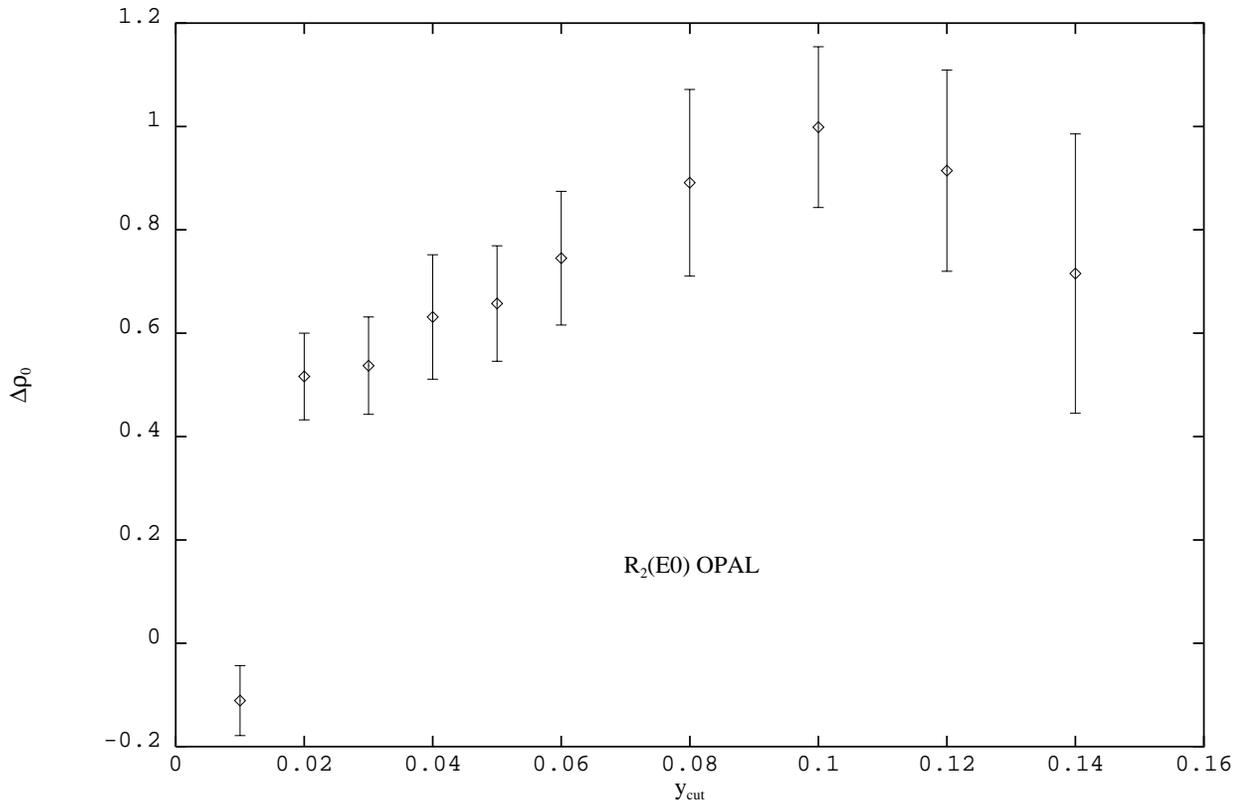
\end{document}